\documentclass[sn-basic]{sn-jnl}

\usepackage{comment}
\usepackage{graphicx}%
\usepackage{amsmath,amssymb,amsfonts}%
\usepackage{amsthm}%
\usepackage{mathrsfs}%
\usepackage[title]{appendix}%
\usepackage{xcolor}%
\usepackage{textcomp}%
\usepackage{manyfoot}%
\usepackage{booktabs}%
\usepackage{algorithm}
\usepackage{algorithmicx}%
\usepackage{algpseudocode}%
\usepackage{listings}%
\usepackage{placeins}
\usepackage{amsmath}
\usepackage{amssymb}
\usepackage{mathtools}
\usepackage[center,font=footnotesize]{caption}
\usepackage{amsthm}
\usepackage[capitalize]{cleveref}

\theoremstyle{definition}

\newcommand{\doi}[1]{\href{https://doi.org/#1}{doi: #1}}

\usepackage{namedtensor}

\allowdisplaybreaks

\ndef{\ax}{ax}
\ndef{\dd}{d}
\ndef{\layer}{layer}
\ndef{\seq}{seq}
\ndef{\subseq}{subseq}
\ndef{\key}{key} \ndef{\val}{val}
\ndef{\heads}{heads}
\ndef{\batch}{batch}
\ndef{\inp}{input} \ndef{\hidden}{hidden} \ndef{\out}{out}
\ndef{\height}{height} \ndef{\width}{width} \ndef{\chans}{chans}
\ndef{\kernel}{kernel} \ndef{\kh}{kh} \ndef{\kw}{kw}
\ndef{\vocab}{vocab}
\ndef{\classes}{classes}

\ndef{\emb}{emb}

\usepackage{multirow}

\usepackage[
  automark,
  autooneside=false,
  headsepline
]{scrlayer-scrpage}
\clearpairofpagestyles
\ihead{\scriptsize Preprint In Review Aug. 2023}
\ohead{\scriptsize Aakash Bansal, Chia-Yi Su, and Collin McMillan}

\begin {document}

\title[Revisiting File Context for Source Code Summarization]{Revisiting File Context for Source Code Summarization}


\author[1]{\fnm{Aakash} \sur{Bansal}}\email{abansal1@nd.edu}

\author[1]{\fnm{Chia-Yi} \sur{Su}}\email{csu3@nd.edu}

\author[1]{\fnm{Collin} \sur{McMillan}}\email{cmc@nd.edu}

\affil*[1]{\orgdiv{Department of Computer Science}, \orgname{University of Notre Dame}, \orgaddress{\street{Holy~Cross Dr}, \city{Notre Dame}, \postcode{46556}, \state{Indiana}, \country{USA}}}

\abstract{Source code summarization is the task of writing natural language descriptions of source code.  A typical use case is generating short summaries of subroutines for use in API documentation.  The heart of almost all current research into code summarization is the encoder-decoder neural architecture, and the encoder input is almost always a single subroutine or other short code snippet.  The problem with this setup is that the information needed to describe the code is often not present in the code itself -- that information often resides in other nearby code.  In this paper, we revisit the idea of ``file context'' for code summarization.  File context is the idea of encoding select information from other subroutines in the same file.  We propose a novel modification of the Transformer architecture that is purpose-built to encode file context and demonstrate its improvement over several baselines.  We find that file context helps on a subset of challenging examples where traditional approaches struggle.}




\maketitle

\section{Introduction}

A source code ``summary'' is a short description of that code in natural language. Code summaries have long been at the center of documentation for programmers such as JavaDocs and PyDocs~\citep{kramer1999api}, though recent interest is ballooning for interactive programmer tools and educational systems.  Tools with code summarization features such as Github Copilot and ChatGPT have captured the public imagination with their ability to read and describe code. Even a short summary such as ``books a seat on an airplane flight'' can help a programmer quickly understand what a snippet of source code does without having to read the code itself.

The beating heart of almost all source code summarization research is the neural encoder-decoder architecture~\citep{sutskever2014sequence}. The setup is simple. The input to the encoder is the source code to be summarized, while the decoder generates a summary one word at a time. Usually in laboratory settings, the code to be summarized is a single subroutine. The problem with this setup is that not all of the information needed to write a summary of a subroutine is included in that subroutine. Software engineering literature has documented for decades how programmers need summaries to include high-level rationale about why the code exists rather than just restating words from the code itself~\citep{holmes2005using, hill2009automatically}. Current approaches struggle to find the right words to provide this rationale because it often does not exist in single subroutine.

An alternative model was proposed by~\citet{haque2020improved} using ``file context.''  File context means the other subroutines in the same file as a subroutine under investigation. Haque~\emph{et al.} built an encoder based on recurrent neural networks (RNN) to augment the attention component of the RNN-based encoder-decoder architectures that were prevalent at the time.  Experimental results in the original paper and replicated by~\citet{bansal2021project} showed how the file context encoder could be added to and improves several RNN-based baselines.

Since then Transformer-based architectures of superseded RNN-based ones in nearly all respects.  Transformer models tended to be able to outperform RNNs even with the file context encoder, when performance is measured by automated metrics over a whole dataset.  E.g., BLEU scores were higher over popular benchmark datasets.  Yet these results are slightly misleading.  In fact, there is a subset of code for which file context helps, and a subset where it does not.  Newer models achieve higher BLEU score by boosting performance on part of the dataset, but not in the way in which file context can help.  And unfortunately, there is not a clear means to augment Transformer-based models with the encoder proposed by~\citet{haque2020improved} due to the differences in how Transformer and RNN-based architectures handle attention.

In this paper, we introduce a novel modification to the Transformer architecture for code summarization designed to encode file context. In a nutshell, our approach has two Transformer-based encoders: one for file context and one for the code being described.  On the decoder side we use a stack two Transformer-based decoders.  The ``lower'' decoder receives the output of the file context encoder.  The ``upper'' decoder receives the output of the first decoder and the code encoder. This dual encoder design is an alternative to standard transformers that use a giant context window.

We show experimentally that our approach outperforms several baselines over three datasets, two in Java and one in Python. We also show how the design decision of a separate encoder for file context helps performance. We show that our approach outperforms other transformer architecture that use a large context window or ``prompt''. But more importantly, we show that the performance increase is due to the file context instead of other factors such as scale. We also conduct a human study and report programmer opinions on the quality of summaries generated by our approach, when compared against the best performing baseline.

\section{File Context}
\label{sec:filecontext}

``File context'' is a term in Software Engineering research literature that means the other information in the same file as a section of code under investigation~\citep{holmes2005using, hill2009automatically, guerrouj2014experimental, ding2022cocomic}.  In this paper, as in the earlier work by~\citet{haque2020improved}, the sections of code under investigation are subroutines, and the file context includes a few of the other subroutines in the same file.  File context has been cited for decades as a key source of information for understanding source code, since code lives in an ecosystem of interdependent software components.

\begin{figure}[h!]
\vspace{-0.5cm}
	\centering
	\setcounter{figure}{0}
	\renewcommand{\figurename}{Example}
	\hspace{0.3cm}
	{
		\small
		\begin{verbatim}
			         public void setIntermediate(String intermediate)
			         {  this.intermediate = intermediate;  }
		\end{verbatim}
	}
	\vspace{0.3cm}
	{	
		\begin{tabular}{ll} 
			\emph{reference} & sets the intermediate value for this flight  	\\ \hline
			\citet{leclair2019neural}    & sets the intermediate value for this \textless UNK\textgreater		\\ 
			\citet{haque2020improved} & sets the intermediate value for this flight 		\\ 
			
		\end{tabular}
	}
	\vspace{0.5cm}
	{
		
	\hspace*{-2cm}\begin{tabular}{l}
	1. public void set airline name string airline     \\
	2. public void set destination string destination          \\
	3. public long get flight id return flight id           \\
	4. public void set flight id long flight id this flight id                              \\
	5. public void set flight number string flight number  \\
	\end{tabular}
	\vspace{0.1cm}
	\caption{(upper) Source code for method ID 26052502 in the {\small\texttt{java-long}} dataset.  (mid) Reference summary and summaries generated by an RNN baseline and the same baseline enhanced with file context.  (lower) The file context.}
		
	}
 \vspace{-0.5cm}
\end{figure}

Consider Example~1 from \citet{haque2020improved}.  The Java method {\small\texttt{setIntermediate()}} is a simple setter type function.  Most baselines are capable of writing a summary to this effect, such as the example from ~\citet{leclair2019neural} shown.  But this summary is not that useful by itself because even a novice programmer is likely to expend almost no effort reading the code.  What is useful is to know is \emph{why} the value is set~\citep{roehm2012professional}.  

In Example~1, this why-information is only evident when we consider the file context. Note the file context consists of terms such as ``airline'', ``flight'', and ``destination'' in the signature of other methods. The model using file context was able to find and learn to use the word ``flight'' correctly, thus predicting a more useful summary for the method.

Cases like these are very common in code summarization samples and have a strong impact on model performance. Consider the distribution of our dataset presented in Table~\ref{tab:pie}. The term $wo$ refers to the number of words that are in both the reference summary and file context, but are not present in the method itself:
\vspace{-0.2cm}
\begin{align}
wo = |(FCW - MW) \cap SW|
\end{align}

Here, FCW is the set of words in the file context, MW is the set of words in the method, and SW is the set of words in the summary.  Note that only around 35\% of methods have $wo=0$, a majority contain at least one word in the file context that is not present in the method itself.

\begin{table}[t!]
	\centering
	\normalsize
	\caption{Model performance declines as word overlap between the code summary and the file context increases.  The first column shows $wo$ which is the number of words in the summary that are in the file context but not in the subroutine.  The second column shows METEOR scores for a typical Transformer baseline.\vspace{1.5mm}}
	\begin{tabular}{llll}
		Word Overlap    & METEOR &    \multicolumn{2}{l}{\hspace{0.5cm}Portion of Dataset} \\
		wo = 0               & 41.44  & \includegraphics{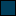} & \multirow{5}{*}{\includegraphics[height=3.7cm]{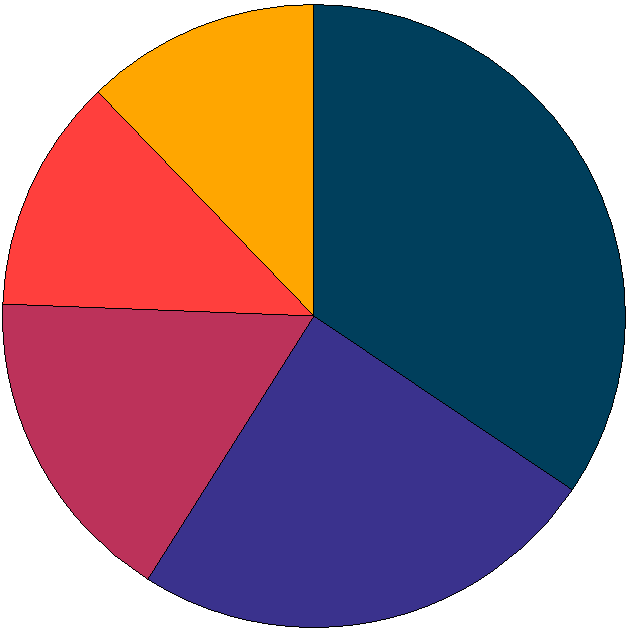}} \\
		\\
		wo = 1               & 30.78  & \includegraphics{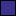} &                   \\
		\\
		wo = 2               & 29.26  & \includegraphics{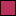} &                   \\
		\\
		wo = 3               & 28.65  & \includegraphics{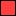} &                   \\
		\\
		wo \textgreater{}= 4 & 21.88  & \includegraphics{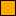} &         \\    
	\end{tabular}
	\label{tab:pie}
\end{table}

Meanwhile, the column METEOR refers to the METEOR score of a Transformer approach proposed by~\citet{ahmad2020transformer}. This baseline approach does not see or use file context. The METEOR score is 41.44 for when $wo=0$, but drops over 25\% to 30.78 for when $wo=1$. We observe that the METEOR score continues to decline for subsets with a higher $wo$. This means summaries that use more words from the file context are often more difficult to write for standard approaches.

\section{Related Work}
\label{sec:related}

In Figure~\ref{fig:related} we provide an overview of selected related work from the last five years. The related work can be broadly classified into four families.

\textbf{AST-Flat} is a family of approaches marked by Column A in Figure~\ref{fig:related} that uses the Abstract Syntax Tree (AST) as a sequence of nodes. In 2018,~\citet{hu2018deep} introduced DeepCom, a model that encodes code tokens and AST nodes together using LSTMs. In 2019,~\citet{leclair2019neural} introduced a similar model ast-attendgru, except that they use GRUs instead of LSTM and they decoupled the AST and source code sequence to separate encoders. They found that learning separate representations of code and structure helps generate better natural language summaries.

\textbf{AST-GNN} is a family of approaches marked by Column G in Figure~\ref{fig:related} that use Graph Neural Networks (GNN). In 2020,~\citet{leclair2020improved} introduced a hybrid GNN-RNN model that encodes AST as a graph using a GNN layer and combines it with a GRU based source code sequence encoder. They found that, compared to a flat representation of AST, a GNN can learn to better place AST nodes in the embedding space by using the edges of the AST. Since that landmark paper, a few approaches have been introduced that use GNNs to encode code structure for code summarization~\citep{kuang2022code,liu2021retrievalaugmented} .

\begin{figure*}[ht]
    \centering
    {
		\begin{tabular}{p{7cm}ccccc}
			 &R          &A          &G      &T      &C                \\
			\citet{hu2018summarizing}					& x	&   &   &  & \\
			\citet{hu2018deep}							& x	&x   &   &  & \\
			\citet{wan2018improving}					& x	&   &   &  & \\
			\citet{liang2018automatic}					& x	&   &   &  & \\
			\citet{alon2019code2vec}					& x	&x   &   &  & \\
			\citet{leclair2019neural}					& x	&x   &   &  & \\
			\citet{nie2019framework}					& x	&   &   &  & \\
			\citet{haldar2020multi}					& x	&x   &   &  & \\
			\citet{ahmad2020transformer}				&  	&   &   &x &  \\
			\citet{haque2020improved}			& x	&   &  & &x  \\ 
			\citet{leclair2020improved}			& x	&x &x    & & \\
			\citet{feng2020codebert}					& 	&   &  &x  &\\
			\citet{wei2020retrieve}					&x 	&x   &  &  &x\\
			\citet{bansal2021project}					& x	&   & & &x  \\ 
			\citet{zugner2021languageagnostic}		& x	&x   &   & &  \\
			\citet{ahmad2021unified}					&	&   &   &x &  \\
			\citet{liu2021retrievalaugmented}				& 	&x   &x  & &x   \\
           	 	\citet{li2022setransformer}                 &  &x   &  &x & \\
			\citet{kuang2022code}			&	&x	&x	&x	& \\
          		\citet{tang2022ast}                        &  &x  &  &x & \\
			\citet{ahmed2022few}		&  &  &  &x & \\
			\emph{this paper}				&	&	&	&x 	&x\\
		\end{tabular}
        \vspace{0.2cm}
    }
    \caption{Overview of select related work. Column R denotes use of RNNs such as GRU/LSTM. A denotes using Abstract Syntax Tree. G denotes graph neural networks based encoder. T denotes Transformer based encoder. C denotes the use of context. }
    \label{fig:related}
    \vspace{-0.8cm}
\end{figure*}

\textbf{Transformers} is a family of approaches marked by Column T in Figure~\ref{fig:related} that mainly use a Transformer architecture. In 2020,~\citet{ahmad2020transformer} introduced a Transformer based approach for code summarization. They found that the self-attention mechanism helps transformers better map words in the encoder to words in the decoder. Since then, several approaches have been introduced that use a network of transformers~\citep{ahmad2021unified,tang2022ast,kuang2022code}. In 2022,~\citet{li2022setransformer} introduced SeTransformer that encodes decoupled AST and source code, using a network of transformers and CNNs.

\textbf{Context} is a family of approaches marked by Column C in Figure~\ref{fig:related} that mainly relies on contextual information, i.e., information outside the target function. In 2020,~\citet{haque2020improved} introduced an encoder ``FC'', that encodes file context, i.e., the summaries around the target function in the same file as a separate encoder. They add this encoder to several RNN based baselines. They found that sometimes the important information needed to generate accurate summaries is not present in the target function, but can be found in the functions around it in the same file. Our paper is an extension of that work.

In 2021, ~\citet{bansal2021project} introduced another context-based encoder ``PC'' that encodes several files from the project as project context. They disclosed that the computational cost of their encoder is exponentially higher than ``FC''. Therefore, we do not replicate their approach for this paper. 

Since 2020, Large Language Models (LLMs) have become increasingly popular in several domains of applied NLP research. In 2020,~\citet{feng2020codebert} introduced CodeBERT that uses a stack of Transformers which are bidirectionally trained. The network models source code syntax by learning to predict randomly masked tokens proposed. In 2021,~\citet{ahmad2021unified} proposed PLBART, using graph neural networks to learn programming language generation. Both report modest improvements in performance for code summarization, while reporting high training costs, which are critical for academic research.

In 2020,~\citet{wei2020retrieve} introduced an approach to use AST and exemplar representations of source code as context. They used these representations to fetch similar methods from a database to help refine the search for a more accurate summary. In 2021,~\citet{liu2021retrievalaugmented} introduced an approach to retrieve similar code properties from a data as context, then model that context as a property graph using GNNs. These are examples of retrieval-based techniques that use summaries from similar methods as an input to the model. In contrast, we make every effort to remove comments and doc-strings from our context input. Retrieval-based techniques are also more susceptible to data leaks between the training and test set that could skew the results over unseen samples.  Therefore, we do not replicate their work as a baseline.

Overall, we observe a clear trend  from GRU/LSTM based models that mainly relied on source code and AST -- to transformer based techniques, with a few approaches increasingly incorporating some contextual information in their design. This paper is a natural evolution of that trend, in that we present an in-depth analysis of how transformers and file context can be used to improve the state-of-the-art in source code summarization.

\vspace{-0.1cm}
\section{Model Design}
\label{sec:model}

Our model is essentially the basic Transformer architecture, but with two key changes.  First, we add a File Context Encoder to learn a representation of the other subroutines in the same file as the target subroutine.  Second, we add another multi-head attention and fully-connected layer unit (which we call an $XFormBlock$) to the decoder, and use the output of the File Context Encoder as an input to this unit. The ``target subroutine'' is the subroutine for which we are writing a summary.  The $XFormBlock$ is an abstraction of common attention, FCN, and normalization/regularization operations used in Transformer-like models (see the next section for more details).  

\begin{figure}[h!]
	\centering
	\includegraphics[width=\linewidth,height=10cm]{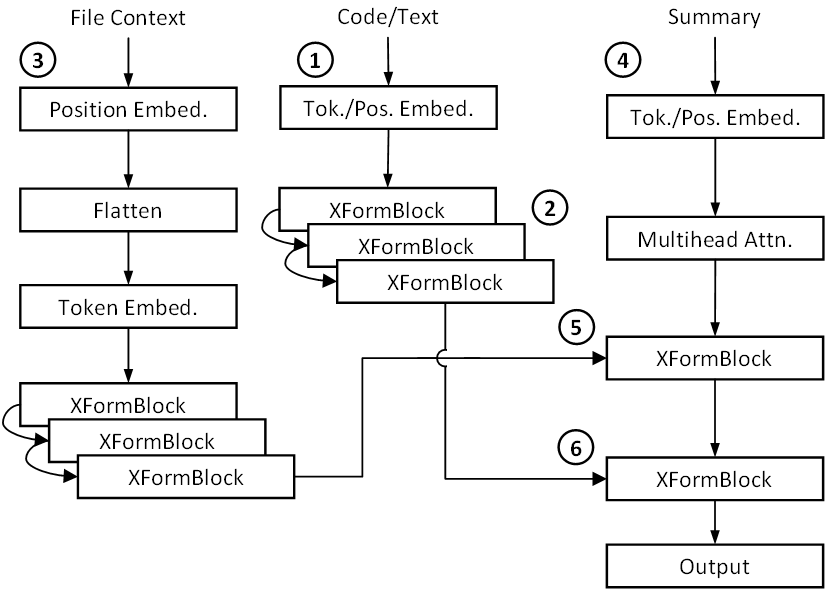}
	\caption{Overview of our model design. Area 1 and 2 denote the target subroutine encoder. Area 3 highlights the novel addition of the file context encoder. Area 4 denotes the summary decoder. Area 5 and 6 denote out novel modifications to use dual encoders to improve the output prediction.}
	\label{fig:overview}
	\vspace{-0.7 cm}
\end{figure}

The intuition behind our changes is two fold: 1) to allow the model to ``see'' the file context prior to the target subroutine's source code when learning to predict words in the summary, and 2) to avoid using giant context windows. Giant context windows, such as large prompts used for LLMs are the industry standard for adding context, in an ad-hoc manner. The expectation is that a large model will eventually learn context from all the information. But, that approach is computationally expensive and requires much a large amount of data. We show Section~\ref{sec:rq3} that giant context window approach does not provide expected improvements, at least not out-of-the-box.

Consider the overview in Figure~\ref{fig:overview}.  There are two encoders and one decoder.  One encoder is the target subroutine encoder, shown in Figure~\ref{fig:overview}, area 1 and 2.  The input to this encoder is the first $t$ tokens of the target subroutine's source code (area~2).  These are encoded using $L$ $XFormBlock$ units, as recommended by~\citet{ahmad2020transformer} for code summarization (area~2).  

Another encoder is the file context encoder shown in area 3.  The input is a matrix containing the first $m$ tokens for $n$ subroutines from the same file.  The vocabulary for both of these encoders is the same, though unlike~\citet{haque2020improved}, we use separate embedding spaces for each encoder.  We used the same preprocessing technique as proposed by~\citet{bansal2021project}, which is very light with only the following operations: 1) extract all tokens based on the language specification (e.g., a ``token'' in Java is what the Java lexical analyzer sees), 2) split all tokens by underscore and camel case, and 3) change all tokens to lower case.

The input to the decoder a sequence of words in the summary for the target subroutine (area 4).  The output is the next word in the summary.  The architecture of the decoder is the same as a typical Transformer model, except that for each $XFormBlock$ in the original model, we have two.  One takes as input the File Context Encoder output (area 5) while another takes the Target Subroutine Encoder's output (area 6).  As in other encoder-decoder architectures, we build the output summary one word at a time by feeding the summary predicted up to that point, back into the decoder.  We use the teacher forcing strategy during training.

In the following subsections, we provide a formal definition of our model in Named Tensor Notation, formalized and proposed by~\citet{chiang2021named}. In our experiments, we call our approach \textbf{\texttt{transformer-fc}}.


\subsection{File Context Encoder}
\label{sec:fcencoder}

The file context encoder forms a representation of the other subroutines in the same file as the target subroutine.  The input to the file context encoder is an $n$ x $m$ matrix where $n$ is the maximum number of other subroutines the encoder can intake and $m$ is the maximum number of tokens per subroutine.  We define the space in which the file context encoder operates as:
\vspace{-0.2cm}
\begin{align}
  \mathbf{X}^{fc} &\in \mathbb{N}^{\name{fcn}[n] \times \name{token}[m]} \notag \\
  \mathbf{P}^{fc} &\in \mathbb{R}^{\name{fcn}[n] \times \name{token}[m] \times \name{dim}[e]} \notag \\
  \mathbf{T}^{fc} &\in \mathbb{R}^{\name{fcn}[n] \times \name{token}[m] \times \name{dim}[e]} \notag \\
  \mathbf{E}^{fc} &\in \mathbb{R}^{\name{fcntoken}[nm] \times \name{dim}[e]} \notag \\
  \mathbf{A} &\in \mathbb{R}^{\name{fcntoken}[nm] \times \name{dim}[e]} \notag
\end{align}
\vspace{-0.6cm}

where $\mathbf{X}^{fc}$ is the input with each element as an entry in the source code vocabulary, $\mathbf{P}^{fc}$ and $\mathbf{T}^{fc}$ are position and token embeddings, and $\mathbf{E}^{fc}$ is the combined embedding space.  $\mathbf{A}$ is the file encoder output space.  We use a learnable position embedding proposed as ``Method 3'' by~\citet{huang2020improve}, in which we assign each position in each subroutine (e.g., $[0..m)$ for each subroutine, rather than $[0..nm)$ across the entire file).  The token embedding is a typical learned space, and $\mathbf{E}^{fc}$ is the elementwise sum of $\mathbf{P}^{fc}$ and $\mathbf{T}^{fc}$, reshaped to a 2d $nm \times e$ matrix from 3d $n \times m \times e$:

\vspace{-0.5cm}
\begin{align}
\mathbf{I}^1 = \left [ 0..n \right )&, \mathbf{J}^1 = \left [ 0..m \right ) \notag \\
\forall i, \forall j, i \in \mathbf{I}^1&, j \in \mathbf{J}^1 \notag
\end{align}
\vspace{-0.6cm}

\vspace{-0.6cm}
\begin{align}
  \mathbf{P}^{fc}_{\name{fcn}(i),\name{token}(j)} &= W^1 \ndot{\name{pos}} \mathbf{J}^1 + b^1  \\
  \mathbf{T}^{fc}_{\name{fcn}(i),\name{token}(j)} &= W^2 \ndot{\name{token}} \mathbf{X}^{fc}_{\name{fcn}(i),\name{token}(j)} + b^2 \notag \\
  \mathbf{E}^{fc}&= (\mathbf{P}^{fc}+\mathbf{T}^{fc})_{(fcn,token)\rightarrow fcntoken} \notag
\end{align}
\vspace{-0.6cm}

\vspace{-0.6cm}
\begin{align}
W^1 &\in \mathbb{R}^{\name{hidden}[e] \times \name{pos[m]}}& b^1 &\in \mathbb{R}^{\name{hidden}[e]} \notag \\
W^2 &\in \mathbb{R}^{\name{hidden}[e] \times \name{token[m]}}& b^2 &\in \mathbb{R}^{\name{hidden}[e]} \notag 
\end{align}
\vspace{-0.6cm}

We apply Transformer-like operations via $L$ layers of self-attention and fully-connected networks, with dropout and layer normalization.  For brevity, we abstract these operations as the function $XFormBlock(Q,K,V)$, as it has been described as an ``Encoder Block'' with parameters Query, Key, and Value by numerous authors starting with~\citet{vaswani2017attention}.

\vspace{-0.5cm}
\begin{align}
\mathbf{A}^{0} &= XFormBlock^0(\mathbf{E}^{fc}, \mathbf{E}^{fc}, \mathbf{E}^{fc})  \\
&\vdotswithin{=} \notag \\
\mathbf{A}^{L} &= XFormBlock^{L-1}(\mathbf{A}^{L-1}, \mathbf{A}^{L-1}, \mathbf{A}^{L-1}) \notag \\
\mathbf{A} &= \mathbf{A}^{L} 
\end{align}

\newpage
\subsection{Target Subroutine Encoder}

The target subroutine encoder forms a representation of the target subroutine itself.  This encoder is a essentially a Transformer-like encoder described by~\citet{vaswani2017attention}, except with the learned position embedding we also use in the file encoder.  The operation space is:

\vspace{-0.6cm}
\begin{align}
\mathbf{X}^{s} &\in \mathbb{N}^{\name{token}[m]} \notag \\
\mathbf{P}^{s} &\in \mathbb{R}^{\name{token}[m] \times \name{dim}[e]} \notag \\
\mathbf{T}^{s} &\in \mathbb{R}^{\name{token}[m] \times \name{dim}[e]} \notag \\
\mathbf{E}^{s} &\in \mathbb{R}^{\name{token}[nm] \times \name{dim}[e]} \notag \\
\mathbf{B} &\in \mathbb{R}^{\name{token}[nm] \times \name{dim}[e]} \notag
\end{align}
\vspace{-0.4cm}

where $\mathbf{X}^{s}$ is the input where each element is an entry in the vocabulary, $\mathbf{P}^{s}$ and $\mathbf{T}^{s}$ are position and token embeddings, and $\mathbf{E}^{s}$ is the combined embedding space:

\vspace{-0.4cm}
\begin{align}
\mathbf{J}^2 = \left [ 0..t \right ) \notag \\
\forall j,  j \in \mathbf{J}^2 \notag
\end{align}
\vspace{-0.4cm}

\vspace{-0.6cm}
\begin{align}
\mathbf{P}^{s}_{\name{token}(j)} &= W^3 \ndot{\name{pos}} \mathbf{J}^2 + b^3  \\
\mathbf{T}^{s}_{\name{token}(j)} &= W^4 \ndot{\name{token}} \mathbf{X}^{s}_{\name{token}(j)} + b^4  \\
\mathbf{E}^{s}&= \mathbf{P}^{s}+\mathbf{T}^{s} 
\end{align}
\vspace{-0.4cm}

\vspace{-0.6cm}
\begin{align}
W^3 &\in \mathbb{R}^{\name{hidden}[e] \times \name{pos[m]}}& b^3 &\in \mathbb{R}^{\name{hidden}[e]} \notag \\
W^4 &\in \mathbb{R}^{\name{hidden}[e] \times \name{token[m]}}& b^4 &\in \mathbb{R}^{\name{hidden}[e]} \notag 
\end{align}
\vspace{-0.4cm}

Followed by $L$ $XFormBlock$ layers:

\vspace{-0.4cm}
\begin{align}
\mathbf{B}^{0} &= XFormBlock^0(\mathbf{E}^{s}, \mathbf{E}^{s}, \mathbf{E}^{s})  \\
&\vdotswithin{=} \notag \\
\mathbf{B}^{L} &= XFormBlock^{L-1}(\mathbf{B}^{L-1}, \mathbf{B}^{L-1}, \mathbf{B}^{L-1}) \notag \\
\mathbf{B} &= \mathbf{B}^{L} 
\end{align}
\vspace{-0.6cm}


\subsection{Decoder}

Our decoder is a Transformer-like decoder, except with \textbf{two} $XFormBlock$ layers for each one in the typical design.  In most Transformer-like decoders, there is a self-attention layer followed by $XFormBlock$ layers.  However in our decoder, we have an $XFormBlock$ that receives $\mathbf{A}$ and another that receives $\mathbf{B}$:

\begin{align}
\mathbf{Y}^{d} &\in \mathbb{N}^{\name{word}[w]} \notag \\
\mathbf{P}^{d} &\in \mathbb{R}^{\name{word}[m] \times \name{dim}[e]} \notag \\
\mathbf{T}^{d} &\in \mathbb{R}^{\name{word}[m] \times \name{dim}[e]} \notag \\
\mathbf{E}^{d} &\in \mathbb{R}^{\name{word}[nm] \times \name{dim}[e]} \notag \\
\mathbf{C} &\in \mathbb{R}^{\name{word}[nm] \times \name{dim}[e]} \notag
\end{align}
\vspace{-0.6cm}

The decoder word embedding space:

\vspace{-0.6cm}
\begin{align}
\mathbf{J}^3 = \left [ 0..w \right ) \notag \\
\forall j,  j \in \mathbf{J}^3 \notag
\end{align}
\vspace{-0.4cm}

\vspace{-0.6cm}
\begin{align}
\mathbf{P}^{d}_{\name{token}(j)} &= W^5 \ndot{\name{pos}} \mathbf{J}^3 + b^5  \\
\mathbf{T}^{d}_{\name{token}(j)} &= W^6 \ndot{\name{word}} \mathbf{X}^{s}_{\name{word}(j)} + b^6  \\
\mathbf{E}^{d}&= \mathbf{P}^{d}+\mathbf{T}^{d} 
\end{align}
\vspace{-0.6cm}

\vspace{-0.6cm}
\begin{align}
W^5 &\in \mathbb{R}^{\name{hidden}[e] \times \name{pos[w]}}& b^5 &\in \mathbb{R}^{\name{hidden}[e]} \notag \\
W^6 &\in \mathbb{R}^{\name{hidden}[e] \times \name{word[w]}}& b^6 &\in \mathbb{R}^{\name{hidden}[e]} \notag 
\end{align}
\vspace{-0.4cm}

And the $XFormBlocks$:

\vspace{-0.4cm}
\begin{align}
\mathbf{C}^{0} &= XFormBlock^0(\mathbf{E}^{t}, \mathbf{E}^{t}, \mathbf{E}^{t}) \\
\mathbf{C}^{1} &= XFormBlock^1(\mathbf{C}^{0}, \mathbf{A}, \mathbf{A}) \\
\mathbf{C}^{2} &= XFormBlock^2(\mathbf{C}^{1}, \mathbf{B}, \mathbf{B}) \\
\mathbf{C} &= \mathbf{C}^{2} \notag
\end{align}
\vspace{-0.6cm}

Followed by an output layer to predict the next word in the summary sequence, where $\mathbf{Y}^{p}$ is the output prediction space:

\vspace{-0.4cm}
\begin{align}
\mathbf{Y}^{p} &\in \mathbb{N}^{\name{vocab}[w]} \notag \\
\mathbf{Y}^{p} &= \sigma(W^7 \ndot{\name{vocab}} \mathbf{C}^{s}_{(word,dim)\rightarrow vocab} + b^7) \notag
\end{align}
\vspace{-0.4cm}

\vspace{-0.4cm}
\begin{align}
W^7 &\in \mathbb{R}^{\name{hidden}[e] \times \name{vocab[m]}}& b^7 &\in \mathbb{R}^{\name{hidden}[e]} \notag 
\end{align}
\vspace{-0.4cm}

In practice, the output is a vector with the length $z$ of the summary vocabulary. We apply a softmax activation to this vector. During prediction, an argmax operation on this vector indicates the index of the predicted word in the vocabulary.

\newpage
\subsection{Hyperparameters}
\label{sec:hyper}
For reproducibility and transparency, we list key model hyperparameters in Table~\ref{tab:hyper}.

\begin{table}[h!]
    \vspace{-0.6cm}
	\caption{ Hyperparameters for our model}
	\label{tab:hyper}
	\vspace{-0.1cm}
	\begin{tabular}{|l|l|l|l|}\hline
		Parameter  &  Description                                     & Java & Python \\ \hline
		$n$ & subroutines in file context & 20 & 20 \\
		$m$ & tokens per subroutine       & 25 & 25 \\
		$t$ & tokens in target subroutine & 50 & 50 \\
		$w$ & words in summary            & 13 & 13 \\
		$v$ & source code vocabulary size~~           & 75k & 100k \\
		$z$ & summary vocabulary size               & 10908 & 11000 \\
		$e$ & embedding dimensions                  & 128 & 128 \\
		$L$ & stacked XFormBlock layers   & 3 & 1 \\
		$h$ & attention heads             & 3 & 3 \\
		$b$ & batch size                            & 8 & 50 \\
		$r$ & learning rate                         & 3e-4 & 3e-4 \\ \hline   
	\end{tabular}
 \vspace{-0.4cm}
\end{table}

We chose $n$, $m$, $t$, and $w$ as recommendations from~\citet{haque2020improved} and ~\citet{bansal2021project}.  We chose $v$ and $z$ as recommended by~\citet{leclair2019recommendations}.  We chose $e$, $L$, $h$, $b$, and $r$ from our own pilot studies and hardware limitations. Note that hyperparameters differ between Java and Python datasets because of semantic differences between the two languages, which require different parameters for comparable performance. Java and Python model performances must also be considered independent of each other.

\vspace{-0.15cm}
\section{Experiment Design}

This section describes our experiment to evaluate our model, including our research questions, methods, datasets, baselines, metrics, hardware and software versions, and threats to validity.

\vspace{-0.3cm}
\subsection{Research Questions}

Our research objective is to determine the degree of difference in performance between our model and baseline models.  To that end, we ask the following Research Questions (RQs):

\begin{description}
	\item[\textbf{RQ1}] What is the difference between our model and baseline models over all subroutines in a dataset, in terms of automated metrics?
	\item[\textbf{RQ2}] What is the difference between our model and baseline models when the summary includes words from the file context?
	\item[\textbf{RQ3}] What is the impact of our dual encoder design on model performance, when compared against a combined input approach?
\end{description}

The rationale behind RQ1 is to be consistent with years of related work. Almost all code summarization papers over the past decade have used automated metrics to measure improvement over baselines. This methodology is popular because any meaningful improvement should be observable over a whole dataset.  

The rationale behind RQ2 is that model changes to add file context may provide more improvement in the subset of the dataset in which words in the summary are only present in the file context. We seek to evaluate our approach, when compared with the baselines, over these subsets of our datasets.

The rationale behind RQ3 is to evaluate the impact of our dual encoder design against a single input approach such as those popularized by Large Language Models (LLMs). State-of-Practice in industry and now research venues, is to use these LLMs, which are designed for instructional and conversational use. These models accept a single input sequence which contains all the information as a user query.

\vspace{-0.3cm}
\subsection{Research Methods}

Our research method is typical of many papers on code summarization and encoder-decoder models such as~\citet{leclair2019neural}:

To answer RQ1, we first train our approach and each of the baselines models independently on different datasets for a maximum of ten epochs using the hyperparameters in Section~\ref{sec:model}. Then, we choose the epoch with the highest validation accuracy. We load the model at that epoch to predict the summaries for the unseen test set, which is 4-5\% of the whole dataset. Then, we compute automated metric scores between predicted summaries and the reference summaries (described in Section~\ref{sec:metrics}).

To answer RQ2, we start with the same predictions we computed for RQ1. While computing metrics, we partition the test set based on the value of $wo$. We then report average metric scores for each subset.

To answer RQ3, we implemented two alternative approaches for modeling file context as described in Section~\ref{sec:alternative}.  We follow the same procedures for evaluation of these models as RQ1, except that we only train {\small \texttt{llama-lora}} for one epoch, which takes roughly 60 hours. We report metric scores for comparison against our approach and a previous approach that uses file context.

\subsection{Datasets}

We use three datasets in this paper.  One is called \texttt{funcom-java} and contains around 2m Java methods.  The dataset originates from~\citet{leclair2019recommendations} and uses a split-by-project configuration to reduce the risk of data duplication.  The version we use is from~\citet{bansal2021project} who apply additional filters to reduce code clones as recommended by~\citet{allamanis2019adverse}. 

The second dataset is called \texttt{funcom-java-long} which was created by~\citet{bansal2023human} and consists of the top 10\% longest methods from \texttt{funcom-java} in terms of number of tokens and implement fixes recommended by \citet{shi2022we}.  We focus on these longer functions due to an observation by~\citet{haque2021action} that many Java methods in the \texttt{funcom-java} dataset are short getters/setters. This observation was corroborated by~\citet{bansal2023human}. The found that longer methods are more challenging while shorter methods are easy to summarize using even the most basic code summarization models. This dataset consists of roughly 190k java methods.

Finally, we create a new dataset we call \texttt{funcom-python} that we extracted from a data dump of 40k Python projects downloaded from Github.  We use the same cleaning and splitting procedures as~\citet{leclair2019recommendations} and~\citet{bansal2021project}. We also favor longer Python functions as recommended by~\citet{haque2021action} and~\cite{bansal2023human}. The final dataset consists of 270k python functions and the corresponding AST sequence, AST graphs, file context, and header summaries. 

\vspace{-0.2cm}
\subsection{Baseline Models}

We compare our model to the following baselines.  We reimplemented all baselines in our own framework to control experimental variables such as software version.  We chose these baselines as they are representative of different families of approaches (see Section~\ref{sec:related}) which have different advantages and disadvantages.

\textbf{transformer-alt} This is a baseline of our own design using the same architecture of our proposed model in Section~\ref{sec:model}.  The only difference is that we substitute the file context input with the subroutine code itself (repeated $n$ times so model size is identical).  More formally, prior to equation (2), we set $\mathbf{X}^{fc} = [{\mathbf{X}^{s}}_{\times n}]$. The reason for this baseline is that our model is larger than the normal {\small \texttt{transformer}}, so it is possible that improvements come from scale and not file context. 

\textbf{ast-attendgru} An approach by~\citet{leclair2019neural} that mainly benefits from a flat representation of the code's Abstract Syntax Tree (AST) generated by the Structure-Based-Traversal (SBT) method.  The model architecture consists of two GRU-based encoders, one for the source code tokens and the other for the AST.

\textbf{ast-attendgru-fc} The original approach by~\citet{haque2020improved} that this paper extends. The model architecture consists of three GRU based encoders -- two for the source code tokens and AST tokens respectively, and a third to encode the file context using a number of GRUs consistent with the number of methods from the file being represented as context.

\textbf{codegnngru} A graph neural network-based approach by~\citet{leclair2020improved}.  This model architecture consists of two encoders, a GRU to encode the flat sequence of the AST graph, and GNN-GRU hybrid encoder to encode a graphical representation of the AST using the edges between AST tokens and source code.

\textbf{transformer} An approach by~\citet{ahmad2020transformer}.  Essentially this approach uses a vanilla Transformer encoder-decoder design. The model architecture consists of a single encoder that uses a positional encoding and two attention heads. 

\textbf{setransformer} A recent approach by~\citet{li2022setransformer} that uses a Transformer-CNN hybrid model to learn representation of the AST. The model architecture consists of two Transformer-based encoders, one for the source code and the other for the AST. Each of these encoders also consists of a CNN layer for feature reduction to reduce computational load.

\vspace{-0.25cm}
\subsection{Alternate Approaches}
\label{sec:alternative}

We also test two alternate approaches for encoding file context as stand-ins for transformers and large language models with a giant context window. We combine the source code of the target subroutine and file context into one input. We use these two baselines exclusively to answer RQ3:

\textbf{transformer-comb} This is a Transformer-based encoder-decoder baseline, where we combine the source code and the file context into a single input. The model architecture is similar to {\small \texttt{transformer}}, scaled up for a much longer input, such that the file context is concatenated to the source code after tokenization.

\textbf{llama-lora} A recent approach by~\citet{hu2021lora} that proposes a framework for fine-tuning LLaMA, an instructional Transformer-Based Large Language Model (LLM) by~\citet{touvron2023llama}. This is a decoder-only instructional and conversational LLM that accepts a single prompt and predicts the most likely words to complete the prompt. We fine tune the 7 billion parameter LLaMA model for one epoch. We create a fine-tuning prompt following the typical strategy by~\citet{hu2021lora}: 
\vspace{-0.1cm}
\begin{enumerate}
\item Instruction: the text ``describe the following function''
\item Input: \textless source code for the target method\textgreater
\item FC: \textless source code of the functions in the file context\textgreater
\item Output: \textless the reference summary sequence\textgreater
\end{enumerate}
\vspace{-0.1cm}
Note, the purpose of these approaches is to compare our model against an off-the-shelf approach by~\citet{hu2021lora} which consists of taking a large model and feeding it all the data in a single input. The {\small \texttt{llama-lora}} model is considerably larger than any of our other baselines. Therefore, we only run this baseline over the smaller datasets, namely {\small \texttt{funcom-java-long}} and {\small \texttt{funcom-python}}. Each of these datasets took roughly 110 hours of training and inference. Estimated duration for training and inference over {\small \texttt{funcom-java}} would be 980 hours ($\sim$ 6 weeks).

\vspace{-0.2cm}
\subsection{Metrics}
\label{sec:metrics}

We use three metrics to compare predicted summaries to their reference in the datasets: 

\textbf{METEOR} is a well-known metric that calculates the harmonic mean of unigram precision and recall, introduced by~\citet{banerjee2005meteor}. We use this metric because~\cite{roy2021reassessing} conducted a study with human experts and found that it better correlates to human judgement compared to BLEU.

\textbf{USE} is an encoder-based semantic similarity metric proposed by~\cite{haque2022semantic}. They conducted a human study and found encoder-based metrics have a relatively high correlation with the judgement of human experts compared to BLEU. 

\textbf{BLEU} is an n-gram based popular metric used to evaluate code summarization by almost all related work over the last decade. We report BLEU to be consistent with related work, though with the caveat that METEOR and USE are now preferred.

\vspace{-0.2cm}
\subsection{Hardware and Software}
\label{sec:hardware}

The hardware we used for training and inference for our approach and baselines: AMD 5900x CPU, 2xTITAN RTX with 24GB VRAM each, and 128GB system memory.  

Software that we used includes CUDA 11.2, Tensorflow 2.9.2, Python 3.10, Pandas 1.4, NLTK 3.6, Ubuntu LTS 22.04.

\vspace{-0.2cm}
\subsection{Threats to Validity}

Like all experiments, this paper carries threats to validity that could change our conclusions under different experimental conditions. There are three main threats we try to mitigate in the design of our experiment:

The first threat lies in the datasets. We attempt to mitigate this risk by using large datasets in two different programming languages extracted from a diverse set of repositories. We also clean and process the dataset using techniques recommended in related work~\citep{leclair2019recommendations,bansal2021project}, in an attempt to mitigate the risk of data leaks and skewed results. 

The second threat lies in the automated metrics we use to evaluate the performance of our approach against the baselines. We attempted to mitigate this risk by reporting three metrics, following recommendations by latest related work.

The third threat lies in the hyperparameters of our model. We chose these hyperparameters based on limited pilot studies and related work, as we do not have resources for a large grid search. In theory, different hyperparameters could alter our results and conclusions. To promote transparency and reproducibility, we report and discuss these hyperparameters in Section~\ref{sec:hyper}.

In additional to these threats, minor variations in performance can be seen due to different hardware and software versions. Therefore, we report the hardware and software versions used for this paper in Section~\ref{sec:hardware}.

\vspace{-0.2cm}
\section{Experiment Results}
\label{sec:results}
In this section we discuss our experimental results for the four research questions.

\vspace{-0.2cm}
\subsection{RQ1: Overall Performance}
Table~\ref{tab:overallmetrics} shows the overall performance for each model and each dataset.  We found that performance for {\small \texttt{transformer-fc}} was around 3\%, 1\%, and 6.5\% higher than the nearest baseline for METEOR, USE, and BLEU, respectively, over the {\small \texttt{funcom-java}} dataset.  The differences were narrower for the {\small \texttt{funcom-java-long}} dataset: 2\%, 0.3\%, 1\%.  Results are mixed in Python, as {\small \texttt{transformer-fc}} had the highest scores for METEOR and BLEU, but not USE.

\vspace{-0.5cm}
\begin{table}[h]
	\caption{Metric scores for the three datasets.  Our model is \texttt{transformer-fc}, while \texttt{transformer-alt} is our model without file context input.}
	\label{tab:overallmetrics}
    \footnotesize
    \centering
    \begin{tabular}{cccccccccc}
		\multicolumn{1}{c|}{}&						 \multicolumn{3}{c|}{\texttt{funcom-java}}                                                                                                           & \multicolumn{3}{c|}{\texttt{funcom-java-long}}  &							\multicolumn{3}{c}{\texttt{funcom-python}}                                                                      \\
		\multicolumn{1}{c|}{model}              & \multicolumn{1}{c}{M} 		&\multicolumn{1}{c}{U} 			& \multicolumn{1}{c|}{B} 			& \multicolumn{1}{c}{M} 	& \multicolumn{1}{c}{U} 		& \multicolumn{1}{c|}{B} 		& \multicolumn{1}{c}{M} 			& \multicolumn{1}{c}{U} 				& \multicolumn{1}{c}{B} \\ \hline
		\multicolumn{1}{l|}{ast-attendgru}      &    \multicolumn{1}{c}{35.30}           &  \multicolumn{1}{l}{52.89}              &  \multicolumn{1}{l|}{18.33}                     & \multicolumn{1}{c}{33.21}                 &  \multicolumn{1}{l}{50.12}             &   \multicolumn{1}{l|}{18.94}  		&     \multicolumn{1}{c}{26.80}               &       \multicolumn{1}{l}{43.75}             &         \multicolumn{1}{l}{16.92}                        \\
		\multicolumn{1}{l|}{ast-attendgru-fc}   & \multicolumn{1}{c}{35.71}              & \multicolumn{1}{l}{52.94}                 & \multicolumn{1}{l|}{18.94}                   & \multicolumn{1}{c}{33.52}                   & \multicolumn{1}{l}{50.48}              &  \multicolumn{1}{l|}{18.91}    	&  \multicolumn{1}{c}{27.72}                 &     \multicolumn{1}{l}{44.93}                 &    \multicolumn{1}{l}{16.82}                                \\
		\multicolumn{1}{l|}{codegnngru}         & \multicolumn{1}{c}{35.82}             & \multicolumn{1}{l}{53.26}                   & \multicolumn{1}{l|}{18.77}                   & \multicolumn{1}{c}{32.98}                   & \multicolumn{1}{l}{49.85}              &  \multicolumn{1}{l|}{18.75}    	&  \multicolumn{1}{c}{26.11}                  &   \multicolumn{1}{l}{42.36}                 & \multicolumn{1}{l}{17.33}                                    \\
		\multicolumn{1}{l|}{transformer}        &  \multicolumn{1}{c}{35.68}              &    \multicolumn{1}{l}{54.03}               &   \multicolumn{1}{l|}{18.29}               & \multicolumn{1}{c}{33.18}                     & \multicolumn{1}{l}{51.27}               & \multicolumn{1}{l|}{18.52}       	&   \multicolumn{1}{c}{26.74}                 & \multicolumn{1}{l}{43.86}                     &  \multicolumn{1}{l}{15.68}                                    \\
		\multicolumn{1}{l|}{setransformer}      & \multicolumn{1}{c}{36.01}              & \multicolumn{1}{l}{53.43}                 & \multicolumn{1}{l|}{18.71}                   & \multicolumn{1}{c}{32.47}                     & \multicolumn{1}{l}{49.60}              & \multicolumn{1}{l|}{18.51}      	&     \multicolumn{1}{c}{27.35}              &  \multicolumn{1}{l}{43.70}                  &  \multicolumn{1}{l}{17.60}                                 \\ \hline
  		\multicolumn{1}{l|}{transformer-alt} & \multicolumn{1}{c}{35.84}                & \multicolumn{1}{l}{53.98}                   & \multicolumn{1}{l|}{18.54}                  & \multicolumn{1}{c}{33.98}                    & \multicolumn{1}{l}{52.62}              &  \multicolumn{1}{l|}{19.67}    	&   \multicolumn{1}{c}{28.47}                 & \multicolumn{1}{l}{\textbf{45.64}}                   &\multicolumn{1}{l}{17.58}                               \\
		\multicolumn{1}{l|}{transformer-fc}     & \multicolumn{1}{c}{\textbf{37.12}} & \multicolumn{1}{l}{\textbf{54.61}}       & \multicolumn{1}{l|}{\textbf{20.18}}      & \multicolumn{1}{c}{\textbf{34.67}}        & \multicolumn{1}{l}{\textbf{52.77}}  & \multicolumn{1}{l|}{\textbf{19.90}}   & \multicolumn{1}{c}{\textbf{28.58}}   & \multicolumn{1}{l}{45.45}             &   \multicolumn{1}{l}{\textbf{18.21}}                       
	\end{tabular}

\end{table}
\vspace{-0.3cm}

We make a few observations in these results.  First, BLEU scores for our approach tend to show more improvement than other scores.  One possible explanation for this difference is BLEU's dependence on exact word matches, while METEOR and USE have mechanisms for reducing this dependence.  It is likely that our model is able to find more exact matches due to the additional information in the file context. Second, the difference between our model and baselines is greatest in {\small \texttt{funcom-java}}.  There are two likely explanations: 1) {\small \texttt{funcom-java}} has around ten times more examples and therefore may be providing more opportunity for {\small \texttt{transformer-fc}} to learn from a more diverse dataset, and 2) that dataset has more short samples, which have less internal context and therefore may benefit more from file context.

Finally, we observe that the overall scores are not as high as other papers report~\citep{li2022setransformer, wei2019code}.  We attribute this observation to our use of the split-by-project dataset design and duplicate removal techniques, which are recommended procedures from related work~\citep{allamanis2019adverse, leclair2019recommendations}, that are unfortunately not used in many papers. The results we report are internally comparable but not comparable against those in other papers.

\begin{table*}[t!]
	\setlength{\tabcolsep}{5pt}
	\centering
	\caption{Raw METEOR and USE scores for different values of $wo$.  We omit BLEU for brevity because it is less favored than other metrics (see Section~\ref{sec:metrics}).  Note overall diminished performance for higher thresholds of $wo$.The term $wo$ means the number of words that are in both the reference summary of a method and its file context, but not in the method itself (see Section~\ref{sec:filecontext})} 
	\label{tab:metricsdiffA}
	\footnotesize
	\begin{tabular}{lrrrrrlllll}
		& \multicolumn{5}{c}{METEOR}                                                                                                                                                                    & \multicolumn{5}{c}{USE}                                                                  \\
		\multicolumn{1}{l|}{model}              & \multicolumn{1}{l}{wo=0} & \multicolumn{1}{l}{wo$\geq$1} & \multicolumn{1}{l}{wo$\geq$2} & \multicolumn{1}{l}{wo$\geq$3} & \multicolumn{1}{l|}{wo$\geq$4} & wo=0 & wo$\geq$1 & wo$\geq$2 & wo$\geq$3 & wo$\geq$4 \\ \hline
		\multicolumn{1}{l|}{ast-attendgru}      &    \multicolumn{1}{l}{39.25}                     &      \multicolumn{1}{l}{33.24}                                 &      \multicolumn{1}{l}{30.67}                                  &        \multicolumn{1}{l}{30.14}                 &  \multicolumn{1}{l|}{25.06}             &   \multicolumn{1}{l}{56.30}   &     \multicolumn{1}{l}{52.01}               &       \multicolumn{1}{l}{49.27}             &         \multicolumn{1}{l}{48.69}           &       \multicolumn{1}{l}{44.34}             \\
		\multicolumn{1}{l|}{ast-attendgru-fc}   & \multicolumn{1}{l}{38.90}     & \multicolumn{1}{l}{34.05}                 & \multicolumn{1}{l}{31.78}                   & \multicolumn{1}{l}{31.32}                   & \multicolumn{1}{l|}{26.25}                   &  \multicolumn{1}{l}{55.74}    &  \multicolumn{1}{l}{52.45}                 &     \multicolumn{1}{l}{49.83}                 &    \multicolumn{1}{l}{49.18}                &   \multicolumn{1}{l}{44.95}                 \\
		\multicolumn{1}{l|}{codegnngru}         & \multicolumn{1}{l}{40.18}     & \multicolumn{1}{l}{35.55}                   & \multicolumn{1}{l}{31.02}                   & \multicolumn{1}{l}{30.54}                   & \multicolumn{1}{l|}{25.50}                   &  \multicolumn{1}{l}{56.99}    &  \multicolumn{1}{l}{52.36}                  &   \multicolumn{1}{l}{49.82}                 & \multicolumn{1}{l}{49.18}                    &     \multicolumn{1}{l}{44.97}                  \\
		\multicolumn{1}{l|}{transformer}        &  \multicolumn{1}{l}{40.15}                        &    \multicolumn{1}{l}{33.35}                                     &   \multicolumn{1}{l}{30.67}                                     & \multicolumn{1}{l}{29.96}                                       & \multicolumn{1}{l|}{24.73}                   & \multicolumn{1}{l}{57.45}       &   \multicolumn{1}{l}{53.22}                 & \multicolumn{1}{l}{50.36}                     &  \multicolumn{1}{l}{49.61}                  & \multicolumn{1}{l}{45.38}                   \\
		\multicolumn{1}{l|}{setransformer}      & \multicolumn{1}{l}{39.94}                    & \multicolumn{1}{l}{33.97}                                  & \multicolumn{1}{l}{31.42}                                  & \multicolumn{1}{l}{30.77}                                    & \multicolumn{1}{l|}{25.83}              & \multicolumn{1}{l}{56.47}      &     \multicolumn{1}{l}{52.65}              &  \multicolumn{1}{l}{50.17}                  &  \multicolumn{1}{l}{49.56}                   &  \multicolumn{1}{l}{45.52}                  \\ \hline
  		\multicolumn{1}{l|}{transformer-alt} & \multicolumn{1}{l}{40.38}                     & \multicolumn{1}{l}{33.48}                                   & \multicolumn{1}{l}{30.92}                                  & \multicolumn{1}{l}{30.40}                                  & \multicolumn{1}{l|}{25.21}              &  \multicolumn{1}{l}{57.35}    &   \multicolumn{1}{l}{52.87}                 & \multicolumn{1}{l}{50.33}                   &\multicolumn{1}{l}{49.74}                    &      \multicolumn{1}{l}{45.49}              \\
		\multicolumn{1}{l|}{transformer-fc}     & \multicolumn{1}{l}{40.37}                    & \multicolumn{1}{l}{35.43}                                  & \multicolumn{1}{l}{33.04}                                  & \multicolumn{1}{l}{32.50}                                  & \multicolumn{1}{l|}{27.19}              & \multicolumn{1}{l}{57.88}     & \multicolumn{1}{l}{53.44}                   &       \multicolumn{1}{l}{51.61}             &   \multicolumn{1}{l}{51.04}                 & \multicolumn{1}{l}{47.01}                  
	\end{tabular}
	\vspace{2mm}
	{\\(a) \texttt{funcom-java} dataset.\vspace{2mm}}	
	\begin{tabular}{lrrrrrlllll}
		\multicolumn{1}{l|}{model}              & \multicolumn{1}{l}{wo=0} & \multicolumn{1}{l}{wo$\geq$1} & \multicolumn{1}{l}{wo$\geq$2} & \multicolumn{1}{l}{wo$\geq$3} & \multicolumn{1}{l|}{wo$\geq$4} & wo=0 & wo$\geq$1 & wo$\geq$2 & wo$\geq$3 & wo$\geq$4 \\ \hline
		\multicolumn{1}{l|}{ast-attendgru}      &           \multicolumn{1}{l}{40.02}               &  \multicolumn{1}{l}{29.63}                                       &  \multicolumn{1}{l}{28.21}                                       &  \multicolumn{1}{l}{27.49} & \multicolumn{1}{l|}{20.50}   & \multicolumn{1}{l}{55.42}     &   \multicolumn{1}{l}{47.33}                 &  \multicolumn{1}{l}{45.73}                     &  \multicolumn{1}{l}{44.76}                   & \multicolumn{1}{l}{39.25}                   \\
		\multicolumn{1}{l|}{ast-attendgru-fc}   & \multicolumn{1}{l}{40.33}     & \multicolumn{1}{l}{29.94}                   & \multicolumn{1}{l}{28.46}                   & \multicolumn{1}{l}{28.11}                   & \multicolumn{1}{l|}{21.45}                   & \multicolumn{1}{l}{55.79}      &            \multicolumn{1}{l}{47.70}         &   \multicolumn{1}{l}{46.15}                 &   \multicolumn{1}{l}{45.39}                   & \multicolumn{1}{l}{39.68}                     \\
		\multicolumn{1}{l|}{codegnngru}         & \multicolumn{1}{l}{39.85}     & \multicolumn{1}{l}{29.38}                   & \multicolumn{1}{l}{27.88}                   & \multicolumn{1}{l}{27.14}                   & \multicolumn{1}{l|}{20.47}                   & \multicolumn{1}{l}{55.13}     &    \multicolumn{1}{l}{47.08}       &  \multicolumn{1}{l}{45.46}                   &     \multicolumn{1}{l}{44.48}                &  \multicolumn{1}{l}{38.78}                   \\
		\multicolumn{1}{l|}{transformer}        &        \multicolumn{1}{l}{39.98}                  &  \multicolumn{1}{l}{29.62}                                      &  \multicolumn{1}{l}{28.24}                                      &  \multicolumn{1}{l}{27.70}                                      & \multicolumn{1}{l|}{20.95}                   &  \multicolumn{1}{l}{56.62}    &  \multicolumn{1}{l}{48.46}                  &  \multicolumn{1}{l}{46.89}                  &   \multicolumn{1}{l}{45.94}                 & \multicolumn{1}{l}{40.48}                   \\
		\multicolumn{1}{l|}{setransformer}      & \multicolumn{1}{l}{39.39}                    & \multicolumn{1}{l}{28.84}                                  & \multicolumn{1}{l}{27.63}                                  &\multicolumn{1}{l}{26.90}                                   & \multicolumn{1}{l|}{20.18}              &   \multicolumn{1}{l}{54.90}   &      \multicolumn{1}{l}{46.81}               &  \multicolumn{1}{l}{45.26}                   & \multicolumn{1}{l}{44.42}                    & \multicolumn{1}{l}{39.07}   \\     \hline          
		\multicolumn{1}{l|}{transformer-alt} & \multicolumn{1}{l}{41.02}                    & \multicolumn{1}{l}{30.29}                                  & \multicolumn{1}{l}{28.96}                                  & \multicolumn{1}{l}{28.33}                                  & \multicolumn{1}{l|}{21.21}              &  \multicolumn{1}{l}{57.73}    &             \multicolumn{1}{l}{49.93}      &               \multicolumn{1}{l}{48.28}     &                \multicolumn{1}{l}{47.28}    & \multicolumn{1}{l}{41.83}                   \\
		\multicolumn{1}{l|}{transformer-fc}     & \multicolumn{1}{l}{41.71}                  & \multicolumn{1}{l}{30.98}                                 & \multicolumn{1}{l}{29.70}                                  & \multicolumn{1}{l}{29.08}                                  & \multicolumn{1}{l|}{21.88}              & \multicolumn{1}{l}{57.88}     &   \multicolumn{1}{l}{50.09}                 & \multicolumn{1}{l}{48.84}                   &    \multicolumn{1}{l}{47.85}                &             \multicolumn{1}{l}{42.38}      
	\end{tabular}
	\vspace{2mm}
	{\\(b) \texttt{funcom-java-long} dataset.\vspace{2mm}}
	
 \begin{tabular}{lrrrrrlllll}
		\multicolumn{1}{l|}{model}              & \multicolumn{1}{l}{wo=0} & \multicolumn{1}{l}{wo$\geq$1} & \multicolumn{1}{l}{wo$\geq$2} & \multicolumn{1}{l}{wo$\geq$3} & \multicolumn{1}{l|}{wo$\geq$4} & wo=0 & wo$\geq$1 & wo$\geq$2 & wo$\geq$3 & wo$\geq$4 \\ \hline
		\multicolumn{1}{l|}{ast-attendgru}      & \multicolumn{1}{l}{26.74}                         &  \multicolumn{1}{l}{27.06}                                      &   \multicolumn{1}{l}{25.33}                                     &               \multicolumn{1}{l}{-}                         & \multicolumn{1}{l|}{-}                   & \multicolumn{1}{l}{43.58}     &       \multicolumn{1}{l}{44.44}             &         \multicolumn{1}{l}{42.08}           &         \multicolumn{1}{l}{-}          &            \multicolumn{1}{l}{-}        \\
		\multicolumn{1}{l|}{ast-attendgru-fc}   & \multicolumn{1}{l}{27.61}     & \multicolumn{1}{l}{28.15}                   & \multicolumn{1}{l}{27.98}                   & \multicolumn{1}{l}{-}                   & \multicolumn{1}{l|}{-}                   & \multicolumn{1}{l}{44.79}      &              \multicolumn{1}{l}{45.52}      &                \multicolumn{1}{l}{45.39}    &   \multicolumn{1}{l}{-}                 &         \multicolumn{1}{l}{-}            \\
		\multicolumn{1}{l|}{codegnngru}         & \multicolumn{1}{l}{26.20}     & \multicolumn{1}{l}{25.78}                   & \multicolumn{1}{l}{24.57}                   & \multicolumn{1}{l}{-}                   & \multicolumn{1}{l|}{-}                   &   42.33   &   42.49                 &   41.85                 &       \multicolumn{1}{l}{-}              &            \multicolumn{1}{l}{-}         \\
		\multicolumn{1}{l|}{transformer}        &        \multicolumn{1}{l}{26.61}                  & \multicolumn{1}{l}{27.25}                                       & \multicolumn{1}{l}{27.36}                                       &  \multicolumn{1}{l}{-}                                      & \multicolumn{1}{l|}{-}                   & \multicolumn{1}{l}{43.62}     &          \multicolumn{1}{l}{44.82}          &         \multicolumn{1}{l}{44.14}           &           \multicolumn{1}{l}{-}         &   \multicolumn{1}{l}{-}                 \\
		\multicolumn{1}{l|}{setransformer}      & \multicolumn{1}{l}{27.17}                    & \multicolumn{1}{l}{28.06}                                  & \multicolumn{1}{l}{27.80}                                  & \multicolumn{1}{l}{-}                                  & \multicolumn{1}{l|}{-}              &  \multicolumn{1}{l}{43.36}     &            \multicolumn{1}{l}{45.08}        &             \multicolumn{1}{l}{44.01}        &             \multicolumn{1}{l}{-}         &      \multicolumn{1}{l}{-}                \\ \hline
		\multicolumn{1}{l|}{transformer-alt} & \multicolumn{1}{l}{28.50}                    & \multicolumn{1}{l}{28.34}                                  & \multicolumn{1}{l}{28.02}                                 & \multicolumn{1}{l}{-}                                  & \multicolumn{1}{l|}{-}              & \multicolumn{1}{l}{45.58}     &  \multicolumn{1}{l}{45.90}                  &  \multicolumn{1}{l}{45.16}                  &   \multicolumn{1}{l}{-}                 &   \multicolumn{1}{l}{-}                \\
		\multicolumn{1}{l|}{transformer-fc}     & \multicolumn{1}{l}{28.49}                    & \multicolumn{1}{l}{28.92}                                  & \multicolumn{1}{l}{28.81}                                  & \multicolumn{1}{l}{-}                                  & \multicolumn{1}{l|}{-}              &  \multicolumn{1}{l}{45.22}    &    \multicolumn{1}{l}{46.37}                &   \multicolumn{1}{l}{44.66}                  &        \multicolumn{1}{l}{-}             &   \multicolumn{1}{l}{-}                 
	\end{tabular}
 \vspace{2mm}
 {\\(b) \texttt{funcom-python} dataset.}
 \vspace{-0.4cm}
\end{table*}

\begin{figure}[t!]
    \vspace{-0.2cm}
	\centering
	\begin{minipage}{.48\textwidth}
 		 \centering
 		 \includegraphics[width=\textwidth]{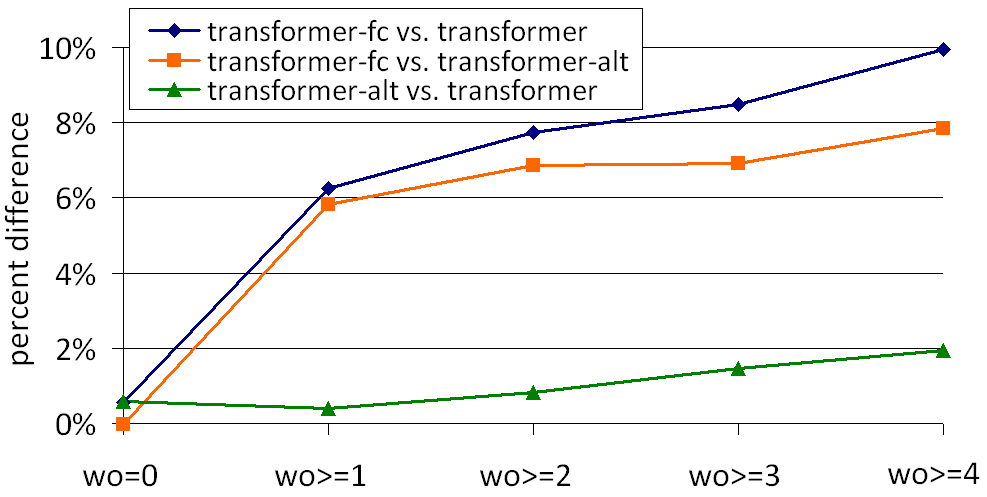}
        \vspace{0.2cm}
  		 \caption{Visual depiction of data for METEOR from Table~\ref{tab:metricsdiffA}(a).  The file context model performance delta increases above 5\% as $wo$ increases, but the delta of the non-file context model does not.}
 		 \label{fig:wometeor}
	\end{minipage}%
	\hfill
	\begin{minipage}{.49\textwidth}
 		\centering
		\vspace{0.2cm}
  		\includegraphics[width=\textwidth]{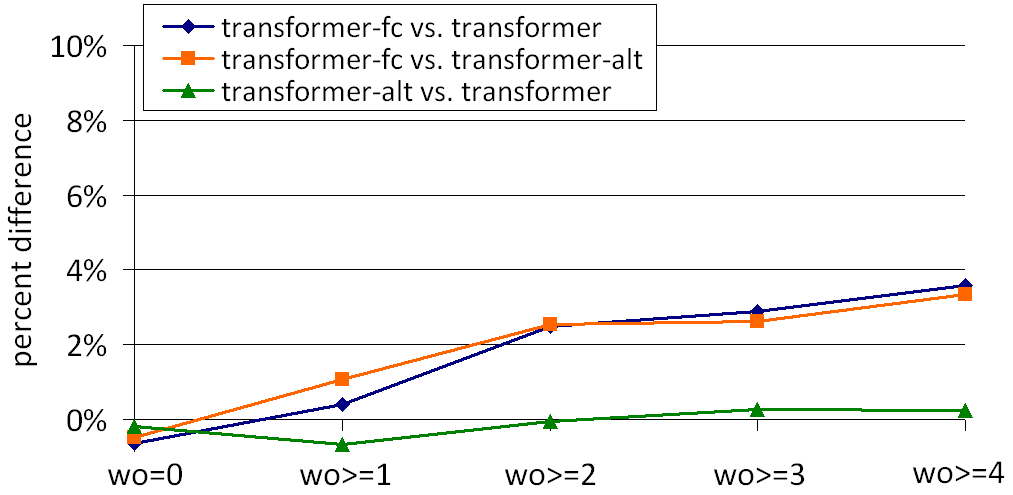}
        \vspace{0.2cm}
  		\caption{Visual depiction of data for USE from Table~\ref{tab:metricsdiffA}(a).  The file context model performance delta decreases when $wo$=0, but increases when $wo\geq1$.  File context is a key factor in overall model improvement.}
  		\label{fig:wouse}
	\end{minipage}
 \vspace{-0.4cm}
\end{figure}
\FloatBarrier

\subsection{RQ2: Effects of File Context}
We report metric scores at different levels of $wo$ in Tables~\ref{tab:metricsdiffA} and~\ref{tab:metricsdiffB}.  We make a few key observations.  First, for the Java datasets, we observe a decline in performance as $wo$ increases across all baselines and all metrics.  We attribute this decline to the difficulty of generating summaries which include ever more information from outside the method being described (see Section~\ref{sec:filecontext}).  This decline is prominent in {\small \texttt{funcom-java-long}}, where METEOR scores when $wo>=4$ tend to be around half compared to $wo=0$.  However, the decline is not consistent in the Python dataset.  METEOR and USE scores decline for {\small \texttt{ast-attendgru}}, {\small \texttt{codegnngru}}, and {\small \texttt{transformer-alt}}, but increase for {\small \texttt{ast-attendgru-fc}} and {\small \texttt{transformer-fc}}.  These results may be expected because the two file context models may improve using file context.

\begin{table*}[t!]
	\centering
    \setlength{\tabcolsep}{5pt}
	\caption{Difference between a given model and a comparison model for each dataset.  Rows indicated for M=METEOR, U=USE, B=BLEU.  For example, the BLEU score for \texttt{transformer-fc} is 27.90\% higher than \texttt{transformer} when $wo>=4$ for the \texttt{funcom-java} dataset. We do not report $wo>=3$ for Python because the number of test samples is very small at those levels (\textless60).}
	\label{tab:metricsdiffB}
    \footnotesize
    \vspace{0.2cm}
	\begin{tabular}{lll|p{0.85cm}p{0.85cm}p{0.85cm}p{0.85cm}p{0.85cm}p{0.85cm}}
		& model              & comparison         & \multicolumn{1}{l}{all} & \multicolumn{1}{l}{wo=0} & \multicolumn{1}{l}{wo$\geq$1} & \multicolumn{1}{l}{wo$\geq$2} & \multicolumn{1}{l}{wo$\geq$3} & \multicolumn{1}{l}{wo$\geq$4} \\ \hline
		\multirow{3}{*}{M} & transformer-fc     & transformer        & 4.04\%                  & 0.55\%                   & 6.24\%                                 & 7.73\%                                 & 8.48\%                                 & 9.95\%                                 \\
		& transformer-fc     & transformer-alt & 3.57\%                  & -0.02\%                  & 5.82\%                                 & 6.86\%                                 &6.91\%                                 & 7.85\%                                 \\
		& transformer-alt & transformer        & 0.45\%                 & 0.57\%                  & 0.39\%                                & 0.82\%                                 & 1.47\%                                 & 1.94\%                                 \\ \hline
		\multirow{3}{*}{U} & transformer-fc     & transformer        & 1.07\%                  & -0.64\%                  & 0.41\%                                 & 2.48\%                                 & 2.88\%                                 & 3.59\%                                 \\
		& transformer-fc     & transformer-alt & 1.17\%                  & -0.47\%                  & 1.08\%                                 & 2.54\%                                 & 2.61\%                                 & 3.34\%                                 \\
		& transformer-alt & transformer        & -0.09\%                 & -0.17\%                  & -0.66\%                                & -0.06\%                                & 0.26\%                                 & 0.24\%                                 \\ \hline
		\multirow{3}{*}{B} & transformer-fc     & transformer        & 10.33\%                 & 2.53\%                   & 15.14\%                                & 16.57\%                                & 16.86\%                                & 27.90\%                                \\
		& transformer-fc     & transformer-alt & 8.85\%                  & 1.48\%                   & 13.31\%                                & 13.73\%                                & 13.28\%                                & 18.85\%                                \\
		& transformer-alt & transformer        & 1.37\%                  & 1.03\%                   & 1.61\%                                 & 2.50\%                                 & 3.17\%                                 & 7.62\%  \\                              
	\end{tabular}
	\vspace{1.5mm}
	{\\(a) \texttt{funcom-java} dataset.\vspace{1.5mm}}
	\begin{tabular}{lll|p{0.85cm}p{0.85cm}p{0.85cm}p{0.85cm}p{0.85cm}p{0.85cm}}
		& model           & comparison      & \multicolumn{1}{l}{all} & \multicolumn{1}{l}{wo=0} & \multicolumn{1}{l}{wo$\geq$1} & \multicolumn{1}{l}{wo$\geq$2} & \multicolumn{1}{l}{wo$\geq$3} & \multicolumn{1}{l}{wo$\geq$4} \\ \hline
		\multirow{3}{*}{M} & transformer-fc  & transformer     & 4.49\%                  & 4.33\%                   & 4.59\%                                 & 5.17\%                                 & 4.98\%                                 & 4.44\%                                 \\
		& transformer-fc  & transformer-alt & 2.03\%                  & 1.68\%                   & 2.28\%                                 & 2.56\%                                 & 2.65\%                                 & 3.16\%                                 \\
		& transformer-alt & transformer     & 2.41\%                  & 2.60\%                   & 2.26\%                                 & 2.55\%                                 & 2.27\%                                 & 1.24\%                                 \\ \hline
		\multirow{3}{*}{U} & transformer-fc  & transformer     & 4.12\%                  & 4.01\%                   & 4.16\%                                 & 4.75\%                                 & 4.50\%                                 & 4.02\%                                 \\
		& transformer-fc  & transformer-alt & 0.29\%                  & 0.26\%                   & 0.32\%                                 & 1.16\%                                 & 1.21\%                                 & 1.31\%                                 \\
		& transformer-alt & transformer     & 2.21\%                  & 2.22\%                   & 2.21\%                                 & 2.56\%                                 & 2.30\%                                 & 1.33\%                                 \\ \hline
		\multirow{3}{*}{B} & transformer-fc  & transformer     & 7.45\%                  & 2.95\%                   & 9.80\%                                 & 9.22\%                                 & 9.24\%                                 & 13.91\%                                \\
		& transformer-fc  & transformer-alt & 1.17\%                  & -2.69\%                  & 3.23\%                                 & 3.40\%                                 & 3.70\%                                 & 9.98\%                                 \\
		& transformer-alt & transformer     & 6.21\%                  & 5.80\%                   & 6.36\%                                 & 5.64\%                                 & 5.34\%                                 & 3.57\% \\                               
	\end{tabular}
    \vspace{1.5mm}
	{\\(b) \texttt{funcom-java-long} dataset.\vspace{1.5mm}}
	\begin{tabular}{lll|p{0.85cm}p{0.85cm}p{0.85cm}p{0.85cm}p{0.85cm}p{0.85cm}}
		& model           & comparison      & \multicolumn{1}{l}{all} & \multicolumn{1}{l}{wo=0} & \multicolumn{1}{l}{wo$\geq$1} & \multicolumn{1}{l}{wo$\geq$2} & \multicolumn{1}{l}{wo$\geq$3} & \multicolumn{1}{l}{wo$\geq$4} \\ \hline
		\multirow{3}{*}{M} & transformer-fc  & transformer     & 3.03\%                 & 8.57\%                  & 6.13\%                                & 5.30\%                                & -                                & -                                \\
		& transformer-fc  & transformer-alt &            0.39\%             &    1.37\%                      &    2.05\%                                    &   2.82\%                                     & -                                      &  -                                      \\
		& transformer-alt & transformer     &           2.63\%              &    7.10\%                      &    4.00\%                                    &  2.41\%                                      &  -                                      & -                                      \\ \hline
		\multirow{3}{*}{U} & transformer-fc  & transformer     &   3.63\%                      &       3.67\%                   &  3.46\%                                      & 1.18\%                                       &  -                                      & -                                      \\
		& transformer-fc  & transformer-alt &             -0.42\%            &    -0.79\%                      & 1.02\%                                       & -1.11\%                                       & -                                       & -                                       \\
		& transformer-alt & transformer     &            4.06\%             &   4.49\%                       &     2.41\%                                   &   2.31\%                                     &   -                                     &  -                                      \\ \hline
		\multirow{3}{*}{B} & transformer-fc  & transformer     &   16.14\%                      &           17.22\%               &      12.88\%                                  &    3.97\%                                    &  -                                      & -                                       \\
		& transformer-fc  & transformer-alt &            3.58\%             &     2.15\%                     &  8.94\%                                      &  13.31\%                                      &  -                                      & -                                       \\
		& transformer-alt & transformer     &            12.12\%             &    14.75\%                      &      3.61\%                                  &     -8.24\%                                   &    -                                    &   -      \\                             
	\end{tabular}
    \vspace{1.5mm}
    {\\(c) \texttt{funcom-python} dataset.}
    \vspace{-0.5cm}
\end{table*}
\FloatBarrier

We report metric scores at different levels of $wo$ in Tables~\ref{tab:metricsdiffA} and~\ref{tab:metricsdiffB}.  We make a few key observations.  First, for the Java datasets, we observe a decline in performance as $wo$ increases across all baselines and all metrics.  We attribute this decline to the difficulty of generating summaries which include ever more information from outside the method being described (see Section~\ref{sec:filecontext}).  This decline is prominent in {\small \texttt{funcom-java-long}}, where METEOR scores when $wo>=4$ tend to be around half compared to $wo=0$.  However, the decline is not consistent in the Python dataset.  METEOR and USE scores decline for {\small \texttt{ast-attendgru}}, {\small \texttt{codegnngru}}, and {\small \texttt{transformer-alt}}, but increase for {\small \texttt{ast-attendgru-fc}} and {\small \texttt{transformer-fc}}.  These results may be expected because the two file context models may improve using file context.

However, {\small \texttt{transformer}} and {\small \texttt{setransformer}} also rise from $wo=0$ to $wo>=1$.  One likely explanation is that Python contains fewer functions with reference explanations where file context is present: while in Java, around 65\% of the methods have $wo>=1$, in Python only around 20\% do.  The number of functions where $wo>=2$ is only 4\% of the dataset, compared to around 12\% in Java.  The number of functions in Python where $wo>=3$ is less than 0.5\% -- only 53 functions in the test set, so low that we do not report metric scores due to possible unreliability.  It is likely that factors other than file context are more important to model performance in Python, perhaps causing underperformance when $wo=0$.  In Java, we note relatively high scores when $wo=0$, likely due to many ``easy'' summaries such as ``records a music file'' for a subroutine {\small \texttt{recordMusicFile()}}.

A second observation is that the models which use file context tend to outperform models without it, and the delta between these models tends to increase as the threshold for $wo$ increases. For example, we are able to replicate the result of Haque~\emph{et al.}~\cite{haque2020improved} in showing that {\small \texttt{ast-attendgru-fc}} improves over {\small \texttt{ast-attendgru}}.  But we especially note that our model {\small \texttt{transformer-fc}} improves over {\small \texttt{transformer}} and {\small \texttt{transformer-alt}}. We present the data from Table~\ref{tab:metricsdiffA} in a graphical format in Figures~\ref{fig:wometeor} and~\ref{fig:wouse}.  These show that the differences in METEOR and USE scores between {\small \texttt{transformer-fc}} and {\small \texttt{transformer}} rise above 5\% when $wo>=1$ in the {\small \texttt{funcom-java}} dataset.  So even though the overall performance improvement from our model is around 3\%, we note that the improvement is concentrated among a small set of especially challenging summaries that primarily benefit from file context.

An alternative interpretation is that the delta only seems larger because the baseline scores are lower as the threshold of $wo$ increases -- a 1 METEOR point improvement is 3.3\% of 30 but 5\% of 20.  However, consider the {\small \texttt{transformer-alt}} scores compared to {\small \texttt{transformer}} (the green lines in Figures~\ref{fig:wometeor} and~\ref{fig:wouse}).  METEOR scores do improve between zero and two percent for METEOR, but are essentially flat for USE.  The {\small \texttt{transformer-alt}} model does not include file context but does have architectural differences over {\small \texttt{transformer}}.  The improvements from the scale of {\small \texttt{transformer-alt}} are spread across all levels of $wo$.  Therefore, the evidence suggests that {\small \texttt{transformer-fc}} improves due to file context when $wo>=1$, and not due to architectural differences or mathematical illusions.

\subsection{RQ3: Alternate approaches to model file context}
\label{sec:rq3}
In Table~\ref{tab:rq3}, we report the metric scores for {\small \texttt{transformer-comb}} and {\small \texttt{llama-lora}}, compared against our approach and a previous GRU based file context baseline {\small \texttt{ast-attendgru-fc}}. We observe that  {\small \texttt{transformer-comb}} achieves scores 23-30\% lower than our approach as well as {\small \texttt{ast-attendgru-fc}} for all three datasets. We posit that combining both method and file context into a single input may not be providing the model with enough information to learn how to place the method in the file context. Therefore, these scores suggest that our model design is better suited for the task of using file context to improve code summarization, when compared to a single input Transformer that is provided with the file context simply appended to the source code.

\begin{table}[t!]
	\centering
	\caption{Metric scores over the three datasets comparing different model designs for incorporating file context.\vspace{0.2cm}}
	\label{tab:rq3}
    \small
	\begin{tabular}{l|llll}
		model                 & METEOR & USE   & BLEU  &  \\ \hline
		ast-attendgru-fc      &    35.71    &   52.94    &  18.94     &  \\
		transformer-comb        &  26.07      &  40.68     &  12.67    & \\ \hline
		transformer-fc (ours) & \textbf{37.12}  & \textbf{54.61} & \textbf{20.18} & 
	\end{tabular}
	\vspace{0.1mm}
    \centering
{(a) \texttt{funcom-java} dataset.}
\vspace{3mm}

	\begin{tabular}{l|llll}
		model                 & METEOR & USE   & BLEU  &  \\ \hline
		ast-attendgru-fc      &   33.52     &   50.48    &   18.91    &  \\
		transformer-comb        & 26.24    &  41.18     &  14.09    & \\
		llama-lora         &20.37       & 38.63      &6.99     & \\ \hline
		transformer-fc (ours) & \textbf{34.67}  & \textbf{52.77} & \textbf{19.90} & 
	\end{tabular}
	\vspace{0.1mm}
 \centering
{(b) \texttt{funcom-java-long} dataset.}
\vspace{3mm}

	\begin{tabular}{l|llll}
		model                 & METEOR & USE   & BLEU  &  \\ \hline
		ast-attendgru-fc      &   27.72     &   44.93    &  16.82     &  \\
		transformer-comb        &  19.86      &  34.93     &   11.15   & \\ \hline
		transformer-fc (ours) & \textbf{28.58}  & \textbf{45.45} & \textbf{18.21} & 
	\end{tabular}
 \vspace{0.1mm}
 \centering
{(c) \texttt{funcom-python} dataset.}
\vspace{-0.3cm}
\end{table}

Recall, we only test {\small \texttt{llama-lora}} on {\small \texttt{funcom-java-long}} and {\small \texttt{funcom-python}} due to high estimated training and inference time over the larger dataset. We observe that for {\small \texttt{funcom-java-long}}, {\small \texttt{llama-lora}} achieves scores 40-65\% lower than our approach as well as {\small \texttt{ast-attendgru-fc}}. We do not report the scores for {\small \texttt{funcom-python}} because the scores were less than 1 point for each metric, which means the model is obviously not learning anything. Upon manual inspection we found that the model was prone to predicting code from the target function as the ``response''. It appears the model learned to simply fetch code from the function and file context to reduce training loss. We think that the poor performance of  {\small \texttt{llama-lora}} is because it is originally pre-trained primarily on conversational English data~\citep{touvron2023llama}. Now it is true that recent work such as the short study by~\citet{ahmed2022few} shows promise using LLMs and few-shot learning for code summarization. However, it may simply be that the data we used to fine-tune the model is not enough for the model to re-adjust the learned conversational word embeddings in favor of programming-language specific word associations. 

Given ever-increasing model and prompt length size, it may seem like the ``obvious'' solution is to simply include the entire file context with the target function.  However, we find that that solution is not effective off-the-shelf. We posit that careful model design and improvements are required, when using decoder-only LLMs to learn from file context for source code summarization. In our view, a likely solution is a novel neural architecture, like {\small \texttt{transformer-fc}} that we propose. Additionally, our model can be scaled up to an arbitrary number of layers and attention heads by adjusting our model parameters $L$ and $h$ (see Section~\ref{sec:model}), just like the original Transformer architecture.

\newpage
\section{Human Study}
In this section we describe parameters and results of our human study. This study adds a qualitative evaluation to complement our quantitative evaluation in the last section.

\subsection{Research Questions}
To design our human study and evaluation, we asked two additional research questions:
\begin{description}
	\item[\textbf{RQ4}] When comparing summaries generated by {\small \texttt{transformer-fc}} and {\small \texttt{transformer-alt}}, which ones do programmers consider better in terms of accuracy, conciseness, completeness, and similarity to reference?
	\item[\textbf{RQ5}] When comparing summaries generated by {\small \texttt{transformer-fc}} and {\small \texttt{transformer-alt}}, which ones do programmers prefer overall?
\end{description}

The rationale behind RQ4 is to compare our approach against the best performing baseline, in terms of the most important qualities of a summary from related work~\cite{sridhara2010towards,bansal2023callcon}. These qualities are accuracy, conciseness, completeness, and similarity to reference. Although automated metrics are the standard for evaluation in related work, programmer opinion is important as an indicator of qualities of a summary which automated metrics may not necessarily represent~\citep{haque2022semantic}.

The rationale behind RQ5 is that programmers may prefer one summary over the other for reasons not formalized by RQ4. Although the qualities we ask subjects to rate in RQ4 are extensively used in related work, there may be other qualities that programmers prefer in summaries.

\subsection{Interface}
\begin{figure}[h]
    \centering
    \includegraphics[width=.95\textwidth]{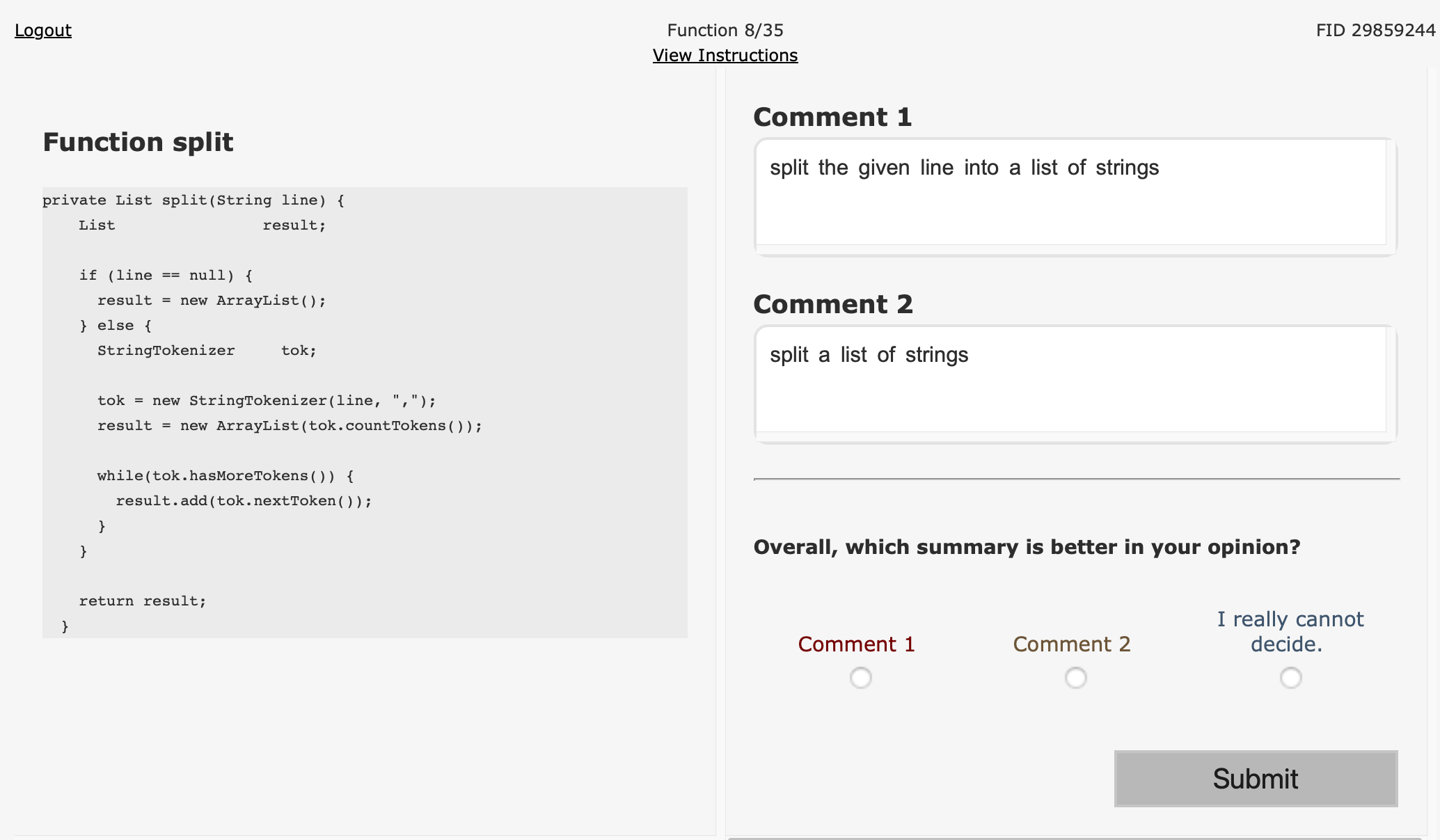}
    \caption{A screenshot of our human study interface}
    \label{fig:interface}
    \vspace{-0.6cm}
\end{figure}

For our human study, we designed a web interface, a screenshot of which is in Figure~\ref{fig:interface}. The interface showed Java source code on the left and two summaries on the top right, comment 1 on top and comment 2 under it. To prevent demand characteristic bias~\citep{dell2012yours}, we do not reveal the source of the comments evaluated. Source of comments 1 and 2 were randomly selected and anonymized for each method and participant. For example, for some methods comment 1 is from our approach {\small \texttt{transformer-fc}} and comment 2 from the baseline transformer-alt, while it is the opposite for other methods seen by the same participant. For each method, participants were asked 5 questions on the bottom right:

\begin{description}
\item[\textbf{Q1}] Independent of other factors, which summary is more accurate?

\item[\textbf{Q2}] Which summary is missing more information that is important for understanding the method?

\item[\textbf{Q3}] Which summary contains more unnecessary information?

\item[\textbf{Q4}] Overall, which summary is better in your opinion?

\item[\textbf{Q5}] Which summary is more similar to this third summary on the left?

\end{description}

For Q5, the participants are shown the reference summary on the bottom left below the code. For each question the participants are presented with three choices: 1)``comment 1'', 2)``comment 2'', and 3)``I really cannot decide''.

\vspace{-0.2cm}
\subsection{Dataset}
The dataset we use for our human study consists of summaries generated for 35 Java methods from the test set of {\small \texttt{funcom-java}}. We select a small subset for human evaluation, because while the automated metrics in Section~\ref{sec:results} are computed over large test sets, human studies are time-restrictive. Extended studies can exhaustion, leading to a decrease in quality and reliability of the data~\citep{jeong2023exhaustive}.

To select these summaries, we filtered the test set for methods where the predicted summaries from {\small \texttt{transformer-fc}} and {\small \texttt{transformer-alt}} differ by at least 2 words. Then, we picked 35 random methods. We restrict the dataset to 35 methods to keep the study duration to around 1 hour, to prevent fatigue bias. The average evaluation time reported by similar studies is $1.5$ minutes/method~\citep{bansal2023callcon}.

\vspace{-0.2cm}
\subsection{Participants}
We recruited 15 Java programmers using Prolific, a web service that facilitates screening and recruitment of research study participants from the UK and USA. We compensated each participant at a flat rate of \$20 for roughly a one hour session. 

\vspace{-0.2cm}
\subsection{Threats to Validity}
Like any human study, the biggest threats to validity are from participant exhaustion or bias. To mitigate the threat of participant exhaustion we restrict the time of our study to around 1 hour as recommended in related work~\citep{sievertsen2016cognitive}. To mitigate the threat of participant bias, we designed our interface as a blind test, without revealing the source of comments. We randomly generate the order of samples shown, which is different for each of the 15 participants. We also analyze their answers in a post-processing step to look for suspicious patterns such as same option for successive choices or identical choices between participants. We did not find any samples that exhibit these patterns.

\section{Human Study Results}
In this section we report and discuss the results of the two additional RQs we asked for the human study.
\subsection{RQ4: Qualitative Comparison}
\begin{figure}[b!]
\vspace{-0.45cm}
    \centering
    \includegraphics[width=0.65\textwidth]{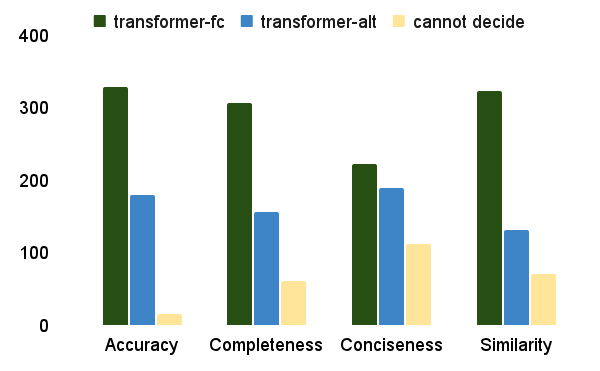}
    \caption{Qualitative comparison of {\small \texttt{transformer-fc}} and {\small \texttt{transformer-alt}}. The participants were also given a third option--cannot decide.}
    \label{fig:human_bar}
\end{figure}

\begin{figure}[t!]
    \centering
    \includegraphics[width=0.9\textwidth]{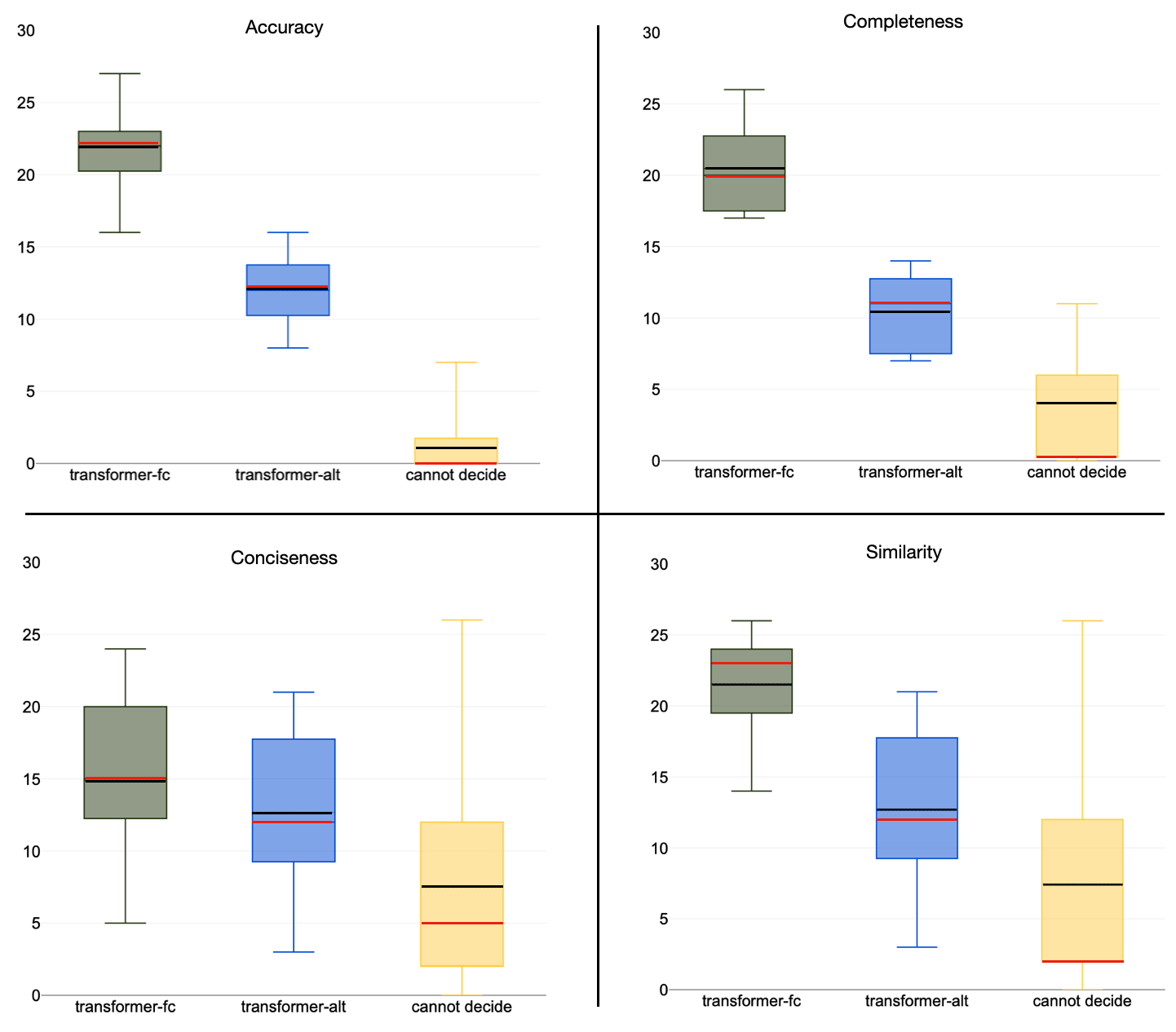}
    \caption{This Box-plot shows distribution for accuracy, completeness, conciseness, and similarity (to reference). The whiskers indicate maximum and minimum values across all 15 participants. The red line in each box indicates median and the black line indicates mean.}
    \label{fig:human_box}
    \vspace{-0.6cm}
\end{figure}

In Figure~\ref{fig:human_bar}, we report the distribution of all human ratings. The total number of human ratings is 525, i.e., all 35 methods rated by each of the 15 participants. Note that we phrased completeness and conciseness questions negatively in the interface. During post-processing, we flip those ratings (except where participants could not decide) to obtain positive scores to compare with other qualities. Additionally we present box-plots showing the distribution of these ratings in Figure~\ref{fig:human_box}. 

In terms of accuracy, we found that 62\% of individual ratings picked summaries generated by our approach {\small \texttt{transformer-fc}} as more accurate than the baseline {\small \texttt{transformer-alt}}. In comparison, only 32\% of the individual ratings picked the baseline as more accurate. This re-affirms our hypothesis that file context helps most for a subset of cases, while overall metric scores might be affected by some cases where it may not improve the summary. In Figure~\ref{fig:human_box}, we observe a small standard deviation indicating general consensus between participants, with high median value of 22 samples out of 35 for {\small \texttt{transformer-fc}}. Overall, we observe that {\small \texttt{transformer-fc}} generates more accurate summaries than {\small \texttt{transformer-alt}} for majority of samples.

In terms of completeness, we see similar trends as accuracy, where 58\% of all ratings indicated that our approach generated more complete summaries. In Figure~\ref{fig:human_box} we observe that each participant indicated that {\small \texttt{transformer-fc}} generated summaries were more complete for at least 17 of the 35 samples, with a median of 20 samples. These values are also seen to be higher than the maximum values for {\small \texttt{transformer-alt}}, where each participant found summaries generated by the baseline to be better in 14 or less samples, with a median of 11. Overall, we find that for a majority of samples, participants favored summaries generated by our approach with file context.

In terms of conciseness, we observe closer aggregate scores of 43\% in favor of {\small \texttt{transformer-fc}}, 37\% in favor of {\small \texttt{transformer-alt}}, and 20\% could not decide. A possible reason for this is that our summaries are limited to 13 words. We posed this question negatively in the study, asking which summary had more unnecessary information. Due to the short length of our summaries, it may have been harder for participants to decide which information was unnecessary. In Figure~\ref{fig:human_box}, we observe a lot of overlap in the distributions. Overall, {\small \texttt{transformer-fc}} achieves a higher median and mean when compared to {\small \texttt{transformer-alt}}, albeit with a smaller margin than other qualities.

In terms of similarity to reference, we observed that the participants found it difficult to decide similarity to reference for 13\% of all samples. One possibility is that even though one of the generated summaries was more accurate, the reference summary may be completely different, such as generic summaries from Javadocs. In Figure~\ref{fig:human_box} we observe that the minimum number of methods in favor of our approach is higher than the mean and median values for the distribution of the baseline. Also, the median value for our approach is higher than the maximum value for the baseline.  Overall, a majority of participants found {\small \texttt{transformer-fc}} generated summaries to be more similar to reference by a considerable margin.

In short, we observe that subjects of our human study found summaries generated by {\small \texttt{transformer-fc}} to be more accurate,complete, and similar to reference when compared to transformer-alt. For conciseness, the results are not as clear, which maybe attributed to the fact that we limit our summaries to a maximum of 13 words.

\subsection{RQ5: Overall Preference}
\begin{figure}[h!]
    \vspace{-0.4cm}
    \centering
    \includegraphics[width=0.8\textwidth]{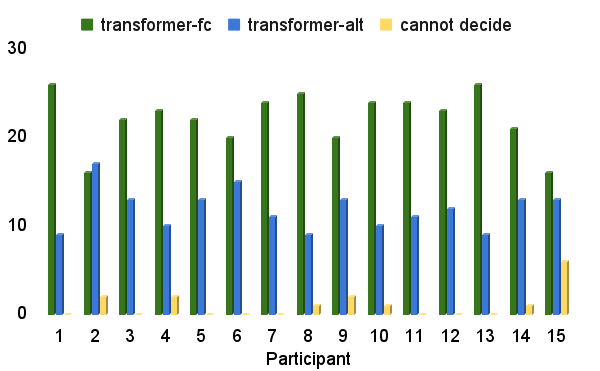}
    \caption{ Overall preference of each participant when presented with summaries generated by {\small \texttt{transformer-fc}} and {\small \texttt{transformer-alt}}. The participants were unaware of the source of the summaries. The y-axis denotes number of samples marked each answer and x-axis denotes the participants.}
    \label{fig:overall}
    \vspace{-0.6cm}
\end{figure}

In Figure~\ref{fig:overall} we report the distribution of overall preference for each of the 15 participants. We observe that 13 out of 15 participants found summaries generated by {\small \texttt{transformer-fc}} to be better overall for a majority of the samples (50\% or more). For participant numbers 2 and 15, our approach did not reach the majority threshold of 18 samples, but neither did the baseline. A few outliers are expected in human studies such as participant 2, but a vast majority of participants favored summaries generated by {\small \texttt{transformer-fc}} when compared with summaries generated by {\small \texttt{transformer-alt}}.

\FloatBarrier

We also performed a Mann-Whitney U test to measure the statistical significance of this difference in our distribution of participants. We computed values of $U1=223$ and $p-val=4.6e^{-6}$. As $p<<<0.05$, we reject the null hypothesis and find that the difference is statistically significance. Overall, participants preferred summaries generated by our approach {\small \texttt{transformer-fc}}, with a statistically significant margin, when compared to summaries generated by the best performing baseline of the same model size without file context, {\small \texttt{transformer-alt}}.

\vspace{-0.2cm}
\section{Conclusion}
\label{sec:conclusion}

This paper advances the state of the art in three ways: First, we present a neural model for source code summarization that augments a standard Transformer encoder/decoder architecture to accept file context. We propose a novel architecture as an alternative to the popular practice of using large context window that rely on model size alone. We evaluate our model against several baselines over three datasets in two programming languages.  We show that our model outperforms these baselines under our experimental conditions according to three metrics from related work.

Second, we demonstrate that file context is a key factor in source code summarization and the improvements gained by our model.  We report model performance at different levels of the overlap between file context and the summaries, for words not appearing in the code being summarized (we denote this value $wo$ according to the formula in Section~\ref{sec:filecontext}).  We find that in the Java datasets, a marked decrease in performance occurs as thresholds of $wo$ increases.  The relationship is less clear in Python, though we still note generally increasing improvement between our approach and the baselines for METEOR and BLEU.  

Third, we we demonstrate that our model design is well-suited to use the file context. We directly compare our model design against the aforementioned large context window approaches. We evaluate two such alternate approaches. One is a single encoder and decoder Transformer-based network. The other is a decoder-only LLM fine-tuned over one of our datasets. We find that these off-the-shelf approaches that simply combines target source code and file context into a giant context window perform considerably worse than our design. 

Fourth, we conduct a human study to add a qualitative aspect to our evaluation. We find that when presented by two different summaries for the same method, a majority of participants favored summaries generated by our approach compared to the best performing baseline. The participants found summaries generated with file context to be more accurate, complete, similar to reference, and better overall in their opinion. We note that we did not see clear consensus on whether our approach generates more concise summaries than the baseline.

\textbf{Reproducibility} To ensure maximum reproducibility of the results, we release all datasets, source code for our models using Tensorflow 2.10, and trained model files via an online appendix (in a Github repository and an official University archival file sharing service for large dataset files).  We provide an online reproducibility guide with step-by-step instructions showing how we produced the our results:

https://github.com/apcl-research/TransformerFC

\section{Acknowledgements}
This work is supported in part by the NSF CCF-2100035 and CCF-2211428. Any opinions, findings, and conclusions expressed herein are the authors’ and do not necessarily reflect those of the sponsors.

\bibliography{bibliography}%

\begin{thebibliography}{47}
\providecommand{\natexlab}[1]{#1}
\providecommand{\url}[1]{{#1}}
\providecommand{\urlprefix}{URL }
\providecommand{\doi}[1]{\url{https://doi.org/#1}}
\providecommand{\eprint}[2][]{\url{#2}}
 \bibcommenthead

\bibitem[{Ahmad et~al(2020)Ahmad, Chakraborty, Ray, and
  Chang}]{ahmad2020transformer}
Ahmad W, Chakraborty S, Ray B, et~al (2020) A transformer-based approach for
  source code summarization. In: Proceedings of the 58th Annual Meeting of the
  Association for Computational Linguistics, pp 4998--5007

\bibitem[{Ahmad et~al(2021)Ahmad, Chakraborty, Ray, and
  Chang}]{ahmad2021unified}
Ahmad WU, Chakraborty S, Ray B, et~al (2021) Unified pre-training for program
  understanding and generation. arXiv preprint arXiv:210306333

\bibitem[{Ahmed and Devanbu(2022)}]{ahmed2022few}
Ahmed T, Devanbu P (2022) Few-shot training llms for project-specific
  code-summarization. arXiv preprint arXiv:220704237

\bibitem[{Allamanis(2019)}]{allamanis2019adverse}
Allamanis M (2019) The adverse effects of code duplication in machine learning
  models of code. In: Proceedings of the 2019 ACM SIGPLAN International
  Symposium on New Ideas, New Paradigms, and Reflections on Programming and
  Software, pp 143--153

\bibitem[{Alon et~al(2019)Alon, Zilberstein, Levy, and
  Yahav}]{alon2019code2vec}
Alon U, Zilberstein M, Levy O, et~al (2019) code2vec: Learning distributed
  representations of code. Proceedings of the ACM on Programming Languages
  3(POPL):1--29

\bibitem[{Banerjee and Lavie(2005)}]{banerjee2005meteor}
Banerjee S, Lavie A (2005) Meteor: An automatic metric for mt evaluation with
  improved correlation with human judgments. In: Proceedings of the acl
  workshop on intrinsic and extrinsic evaluation measures for machine
  translation and/or summarization, pp 65--72

\bibitem[{Bansal et~al(2021)Bansal, Haque, and McMillan}]{bansal2021project}
Bansal A, Haque S, McMillan C (2021) Project-level encoding for neural source
  code summarization of subroutines. In: 2021 IEEE/ACM 29th International
  Conference on Program Comprehension (ICPC), IEEE, pp 253--264

\bibitem[{Bansal et~al(2023{\natexlab{a}})Bansal, Eberhart, Karas, Huang, and
  McMillan}]{bansal2023callcon}
Bansal A, Eberhart Z, Karas Z, et~al (2023{\natexlab{a}}) Function call graph
  context encoding for neural source code summarization. IEEE Transactions on
  Software Engineering pp 1--14. \doi{10.1109/TSE.2023.3279774}

\bibitem[{Bansal et~al(2023{\natexlab{b}})Bansal, Sharif, and
  McMillan}]{bansal2023human}
Bansal A, Sharif B, McMillan C (2023{\natexlab{b}}) Towards modeling human
  attention from eye movements for neutral source code summarization.
  Proceedings of ACM Human-Computer Interaction, Vol 7

\bibitem[{Chiang et~al(2021)Chiang, Rush, and Barak}]{chiang2021named}
Chiang D, Rush AM, Barak B (2021) Named tensor notation. arXiv preprint
  arXiv:210213196

\bibitem[{Dell et~al(2012)Dell, Vaidyanathan, Medhi, Cutrell, and
  Thies}]{dell2012yours}
Dell N, Vaidyanathan V, Medhi I, et~al (2012) " yours is better!" participant
  response bias in hci. In: Proceedings of the sigchi conference on human
  factors in computing systems, pp 1321--1330

\bibitem[{Ding et~al(2022)Ding, Wang, Ahmad, Ramanathan, Nallapati, Bhatia,
  Roth, and Xiang}]{ding2022cocomic}
Ding Y, Wang Z, Ahmad WU, et~al (2022) Cocomic: Code completion by jointly
  modeling in-file and cross-file context. arXiv preprint arXiv:221210007

\bibitem[{Feng et~al(2020)Feng, Guo, Tang, Duan, Feng, Gong, Shou, Qin, Liu,
  Jiang et~al}]{feng2020codebert}
Feng Z, Guo D, Tang D, et~al (2020) Codebert: A pre-trained model for
  programming and natural languages. arXiv preprint arXiv:200208155

\bibitem[{Guerrouj et~al(2014)Guerrouj, Di~Penta, Gu{\'e}h{\'e}neuc, and
  Antoniol}]{guerrouj2014experimental}
Guerrouj L, Di~Penta M, Gu{\'e}h{\'e}neuc YG, et~al (2014) An experimental
  investigation on the effects of context on source code identifiers splitting
  and expansion. Empirical Software Engineering 19:1706--1753

\bibitem[{Haldar et~al(2020)Haldar, Wu, Xiong, and
  Hockenmaier}]{haldar2020multi}
Haldar R, Wu L, Xiong J, et~al (2020) A multi-perspective architecture for
  semantic code search. arXiv preprint arXiv:200506980

\bibitem[{Haque et~al(2020)Haque, LeClair, Wu, and
  McMillan}]{haque2020improved}
Haque S, LeClair A, Wu L, et~al (2020) Improved automatic summarization of
  subroutines via attention to file context. In: Proceedings of the 17th
  International Conference on Mining Software Repositories, pp 300--310

\bibitem[{Haque et~al(2021)Haque, Bansal, Wu, and McMillan}]{haque2021action}
Haque S, Bansal A, Wu L, et~al (2021) Action word prediction for neural source
  code summarization. In: 2021 IEEE International Conference on Software
  Analysis, Evolution and Reengineering (SANER), IEEE, pp 330--341

\bibitem[{Haque et~al(2022)Haque, Eberhart, Bansal, and
  McMillan}]{haque2022semantic}
Haque S, Eberhart Z, Bansal A, et~al (2022) Semantic similarity metrics for
  evaluating source code summarization. In: Proceedings of the 30th IEEE/ACM
  International Conference on Program Comprehension, pp 36--47

\bibitem[{Hill et~al(2009)Hill, Pollock, and
  Vijay-Shanker}]{hill2009automatically}
Hill E, Pollock L, Vijay-Shanker K (2009) Automatically capturing source code
  context of nl-queries for software maintenance and reuse. In: 2009 IEEE 31st
  International Conference on Software Engineering, IEEE, pp 232--242

\bibitem[{Holmes and Murphy(2005)}]{holmes2005using}
Holmes R, Murphy GC (2005) Using structural context to recommend source code
  examples. In: Proceedings of the 27th international conference on Software
  engineering, pp 117--125

\bibitem[{Hu et~al(2021)Hu, Shen, Wallis, Allen-Zhu, Li, Wang, Wang, and
  Chen}]{hu2021lora}
Hu EJ, Shen Y, Wallis P, et~al (2021) Lora: Low-rank adaptation of large
  language models. arXiv preprint arXiv:210609685

\bibitem[{Hu et~al(2018{\natexlab{a}})Hu, Li, Xia, Lo, and Jin}]{hu2018deep}
Hu X, Li G, Xia X, et~al (2018{\natexlab{a}}) Deep code comment generation. In:
  Proceedings of the 26th Conference on Program Comprehension, ACM, pp 200--210

\bibitem[{Hu et~al(2018{\natexlab{b}})Hu, Li, Xia, Lo, Lu, and
  Jin}]{hu2018summarizing}
Hu X, Li G, Xia X, et~al (2018{\natexlab{b}}) Summarizing source code with
  transferred api knowledge. In: Proceedings of the 27th International Joint
  Conference on Artificial Intelligence, AAAI Press, pp 2269--2275

\bibitem[{Huang et~al(2020)Huang, Liang, Xu, and Xiang}]{huang2020improve}
Huang Z, Liang D, Xu P, et~al (2020) Improve transformer models with better
  relative position embeddings. In: Findings of the Association for
  Computational Linguistics: EMNLP 2020, pp 3327--3335

\bibitem[{Jeong et~al(2023)Jeong, Aggarwal, Robinson, Kumar, Spearot, and
  Park}]{jeong2023exhaustive}
Jeong D, Aggarwal S, Robinson J, et~al (2023) Exhaustive or exhausting?
  evidence on respondent fatigue in long surveys. Journal of Development
  Economics 161:102992

\bibitem[{Kramer(1999)}]{kramer1999api}
Kramer D (1999) Api documentation from source code comments: a case study of
  javadoc. In: Proceedings of the 17th annual international conference on
  Computer documentation, pp 147--153

\bibitem[{Kuang et~al(2022)Kuang, Zhou, and Yang}]{kuang2022code}
Kuang L, Zhou C, Yang X (2022) Code comment generation based on graph neural
  network enhanced transformer model for code understanding in open-source
  software ecosystems. Automated Software Engineering 29(2):43

\bibitem[{LeClair and McMillan(2019)}]{leclair2019recommendations}
LeClair A, McMillan C (2019) Recommendations for datasets for source code
  summarization. In: Proceedings of NAACL-HLT, pp 3931--3937

\bibitem[{LeClair et~al(2019)LeClair, Jiang, and McMillan}]{leclair2019neural}
LeClair A, Jiang S, McMillan C (2019) A neural model for generating natural
  language summaries of program subroutines. In: 2019 IEEE/ACM 41st
  International Conference on Software Engineering (ICSE), IEEE, pp 795--806

\bibitem[{LeClair et~al(2020)LeClair, Haque, Wu, and
  McMillan}]{leclair2020improved}
LeClair A, Haque S, Wu L, et~al (2020) Improved code summarization via a graph
  neural network. In: Proceedings of the 28th international conference on
  program comprehension, pp 184--195

\bibitem[{Li et~al(2022)Li, Wu, Peng, Chen, Sun, Liu, and
  Paul}]{li2022setransformer}
Li Z, Wu Y, Peng B, et~al (2022) Setransformer: A transformer-based code
  semantic parser for code comment generation. IEEE Transactions on Reliability

\bibitem[{Liang and Zhu(2018)}]{liang2018automatic}
Liang Y, Zhu KQ (2018) Automatic generation of text descriptive comments for
  code blocks. In: Thirty-Second AAAI Conference on Artificial Intelligence

\bibitem[{Liu et~al(2021)Liu, Chen, Xie, Siow, and
  Liu}]{liu2021retrievalaugmented}
Liu S, Chen Y, Xie X, et~al (2021) Retrieval-augmented generation for code
  summarization via hybrid {\{}gnn{\}}. In: International Conference on
  Learning Representations,
  \urlprefix\url{https://openreview.net/forum?id=zv-typ1gPxA}

\bibitem[{Nie et~al(2019)Nie, Rai, Li, Khurshid, Mooney, and
  Gligoric}]{nie2019framework}
Nie P, Rai R, Li JJ, et~al (2019) A framework for writing trigger-action todo
  comments in executable format. In: Proceedings of the 2019 27th ACM Joint
  Meeting on European Software Engineering Conference and Symposium on the
  Foundations of Software Engineering, ACM, pp 385--396

\bibitem[{Roehm et~al(2012)Roehm, Tiarks, Koschke, and
  Maalej}]{roehm2012professional}
Roehm T, Tiarks R, Koschke R, et~al (2012) How do professional developers
  comprehend software? In: 2012 34th International Conference on Software
  Engineering (ICSE), IEEE, pp 255--265

\bibitem[{Roy et~al(2021)Roy, Fakhoury, and Arnaoudova}]{roy2021reassessing}
Roy D, Fakhoury S, Arnaoudova V (2021) Reassessing automatic evaluation metrics
  for code summarization tasks. In: Proceedings of the 29th ACM Joint Meeting
  on European Software Engineering Conference and Symposium on the Foundations
  of Software Engineering, pp 1105--1116

\bibitem[{Shi et~al(2022)Shi, Mu, Chen, Wang, Wang, Yang, Li, Xia, and
  Wang}]{shi2022we}
Shi L, Mu F, Chen X, et~al (2022) Are we building on the rock? on the
  importance of data preprocessing for code summarization. In: Proceedings of
  the 30th ACM Joint European Software Engineering Conference and Symposium on
  the Foundations of Software Engineering, pp 107--119

\bibitem[{Sievertsen et~al(2016)Sievertsen, Gino, and
  Piovesan}]{sievertsen2016cognitive}
Sievertsen HH, Gino F, Piovesan M (2016) Cognitive fatigue influences
  students’ performance on standardized tests. Proceedings of the National
  Academy of Sciences 113(10):2621--2624. \doi{10.1073/pnas.1516947113}

\bibitem[{Sridhara et~al(2010)Sridhara, Hill, Muppaneni, Pollock, and
  Vijay-Shanker}]{sridhara2010towards}
Sridhara G, Hill E, Muppaneni D, et~al (2010) Towards automatically generating
  summary comments for java methods. In: Proceedings of the 25th IEEE/ACM
  international conference on Automated software engineering, pp 43--52

\bibitem[{Sutskever et~al(2014)Sutskever, Vinyals, and
  Le}]{sutskever2014sequence}
Sutskever I, Vinyals O, Le QV (2014) Sequence to sequence learning with neural
  networks. Advances in neural information processing systems 27

\bibitem[{Tang et~al(2022)Tang, Shen, Li, Ge, Huang, Zhu, and
  Luo}]{tang2022ast}
Tang Z, Shen X, Li C, et~al (2022) Ast-trans: Code summarization with efficient
  tree-structured attention. In: Proceedings of the 44th International
  Conference on Software Engineering, pp 150--162

\bibitem[{Touvron et~al(2023)Touvron, Lavril, Izacard, Martinet, Lachaux,
  Lacroix, Rozi{\`e}re, Goyal, Hambro, Azhar et~al}]{touvron2023llama}
Touvron H, Lavril T, Izacard G, et~al (2023) Llama: Open and efficient
  foundation language models. arXiv preprint arXiv:230213971

\bibitem[{Vaswani et~al(2017)Vaswani, Shazeer, Parmar, Uszkoreit, Jones, Gomez,
  Kaiser, and Polosukhin}]{vaswani2017attention}
Vaswani A, Shazeer N, Parmar N, et~al (2017) Attention is all you need.
  Advances in neural information processing systems 30

\bibitem[{Wan et~al(2018)Wan, Zhao, Yang, Xu, Ying, Wu, and
  Yu}]{wan2018improving}
Wan Y, Zhao Z, Yang M, et~al (2018) Improving automatic source code
  summarization via deep reinforcement learning. In: Proceedings of the 33rd
  ACM/IEEE International Conference on Automated Software Engineering, ACM, pp
  397--407

\bibitem[{Wei et~al(2019)Wei, Li, Xia, Fu, and Jin}]{wei2019code}
Wei B, Li G, Xia X, et~al (2019) Code generation as a dual task of code
  summarization. Advances in neural information processing systems 32

\bibitem[{Wei et~al(2020)Wei, Li, Li, Xia, and Jin}]{wei2020retrieve}
Wei B, Li Y, Li G, et~al (2020) Retrieve and refine: exemplar-based neural
  comment generation. In: Proceedings of the 35th IEEE/ACM International
  Conference on Automated Software Engineering, pp 349--360

\bibitem[{Z{\"u}gner et~al(2021)Z{\"u}gner, Kirschstein, Catasta, Leskovec, and
  G{\"u}nnemann}]{zugner2021languageagnostic}
Z{\"u}gner D, Kirschstein T, Catasta M, et~al (2021) Language-agnostic
  representation learning of source code from structure and context. In:
  International Conference on Learning Representations,
  \urlprefix\url{https://openreview.net/forum?id=Xh5eMZVONGF}

\end{thebibliography}

\end{document}